\documentclass{sig-alternate-05-2015}

\usepackage{balance}  

\usepackage{mathptmx}
\usepackage{graphicx}
\usepackage{times}
\usepackage{amsmath}
\usepackage{amssymb}

\usepackage{amsthm}
\usepackage{subfigure}
\usepackage[usenames,dvipsnames,table]{xcolor}
\usepackage{xspace}
\usepackage{boxedminipage}
\usepackage{algorithmic}
\usepackage[algoruled]{algorithm2e}
\usepackage{comment}
\usepackage{url}
\usepackage{color}
\usepackage{enumitem}
\usepackage{xfrac}
\usepackage{tabularx}
\usepackage{multirow}
\usepackage{booktabs}
\usepackage{array}
\newcolumntype{M}{>{\centering\arraybackslash}m{\dimexpr.5\linewidth-1.5\tabcolsep}}
\usepackage{bigstrut}
\usepackage[usestackEOL]{stackengine}
\usepackage{txfonts}
\usepackage{bibunits}
\usepackage{wrapfig}

\newtheoremstyle{examplestyle}%
  {3pt}
  {3pt}
  {}
  {}
  {\bfseries}
  {.}
  {.5em}
  {\thmname{#1} \thmnumber{#2} \thmnote{\bfseries(#3)}}
\theoremstyle{examplestyle}

\newtheorem{my-definition}{Definition}

{\begin{list}{$\bullet$}{%
	\setlength{\labelsep}{2pt}\setlength{\leftmargin}{0pt}%
	\setlength{\labelwidth}{0pt}%
	\setlength{\listparindent}{0pt}}}
{\end{list}}

\renewcommand{\paragraph}[1]{\vspace{0.1cm}\noindent \textbf{#1.}}

\newcommand{\eg}{e.g.,\xspace}
\newcommand{\ie}{i.e.,\xspace}

\newcommand{\denselist}{\itemsep 0pt\parsep=1pt\partopsep 0pt}

\newcommand{\highlight}[1]{#1}

\newcommand{\header}[1]{\multicolumn{1}{c}{\textbf{\Shortunderstack{#1}}}}
\newcommand{\headergap}[1]{\multicolumn{1}{c}{\addstackgap{\textbf{\Shortunderstack{#1}}}}}

\newcommand{\dataset}{data set\xspace}
\newcommand{\Dataset}{Data set\xspace}
\newcommand{\datasets}{data sets\xspace}

\newcommand{\DataSets}{Data Sets\xspace}
\newcommand{\DataSet}{Data Set\xspace}

\newcommand{\datapolygamy}{\emph{Data Polygamy}\xspace}

\newcommand{\urbannyc}{\emph{NYC Urban}\xspace}
\newcommand{\opendata}{\emph{NYC Open}\xspace}

\newcommand{\mapreduce}{\emph{map-reduce}\xspace}

\newdef{definition}{Definition}

\newcommand{\Sspace}        {{\mathbb S}}
\newcommand{\Rspace}        {{\mathbb R}}
\newcommand{\Dspace}        {{\mathbb D}}
\newcommand{\Tspace}        {{\mathbb T}}

\usepackage{eqparbox}

\newlength{\oldtextfloatsep}
\newlength{\oldcolumnsep}

\begin{document}


\title{Data Polygamy: The Many-Many Relationships among Urban Spatio-Temporal Data Sets}
%
%
%
%
%

\numberofauthors{1}

\author{ 
\alignauthor Fernando Chirigati{\large$^\ast$}, Harish Doraiswamy{\large$^\ast$}, Theodoros Damoulas{\large$^{\star\dagger}$}, Juliana Freire{\large$^\ast$} \\
\affaddr{{\large$^\ast$} New York University \hspace{1cm} {\large$^\star$} University of Warwick \hspace{1cm} 
{\large$^\dagger$} Alan Turing Institute} \\ \vspace{0.1cm}
\email{\{fchirigati,harishd,juliana.freire\}@nyu.edu \hspace{1cm} damoulas@warwick.ac.uk}
}

\maketitle
\begin{abstract}
The increasing ability to collect data from urban environments, coupled 
with a push towards openness by governments, has resulted in the availability of
numerous spatio-temporal data sets covering diverse aspects of a city. 
Discovering relationships between these data sets can produce new
insights by enabling domain experts to not only test but also
generate hypotheses.
However, discovering these relationships is difficult. First, a relationship
between two data sets may occur only at certain locations and/or time periods. 
Second, the sheer number and size of the data sets, coupled with the diverse spatial 
and temporal scales at which the data is available, presents computational challenges 
on all fronts, from indexing and querying to analyzing them. Finally, it is non-trivial to
differentiate between meaningful and spurious relationships.
To address these challenges, we propose \emph{Data Polygamy}, a
scalable topology-based framework that allows users to query for
statistically significant relationships between spatio-temporal data
sets.
We have performed an experimental evaluation using over 300 spatial-temporal
urban data sets which shows that our approach is scalable and
effective at identifying interesting relationships. 
\end{abstract}

\begin{bibunit}[abbrv]
\section{Introduction}
\label{sec:intro}
Urban environments are the loci of economic activity and innovation.
At the same time, most cities face huge challenges around
transportation, resource consumption, housing affordability, and
inadequate or aging infrastructure.
The growing volumes of urban data being collected and made
available~\cite{barbosa@bigdata2014,opendata@book2013,chicagoopendata,nycopendata,OpenGovDataUK2012}
open up new opportunities for city governments and social scientists
to engage in data-driven science to better understand cities, make
them more efficient, and improve the lives of their residents.

\begin{figure}[t]
\centering
\includegraphics[width=0.95\linewidth]{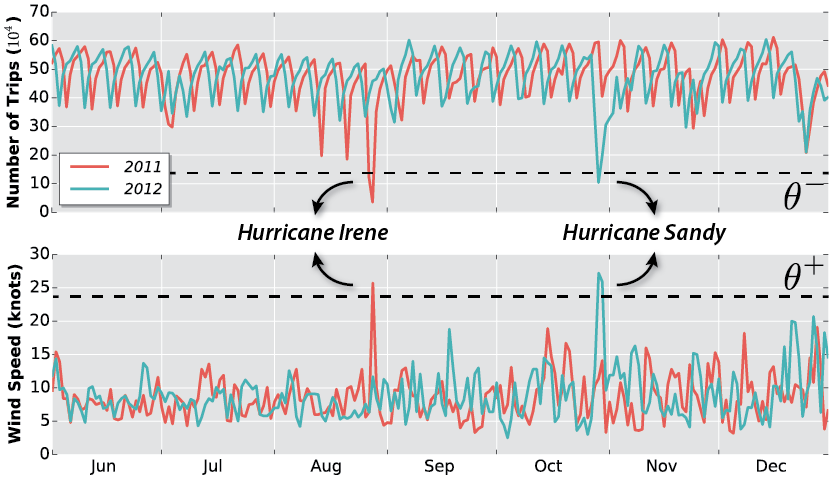}
\vspace{-0.15in}
\caption{Variation of the number of taxi trips in NYC and its
  relationship with wind speed.}
\label{fig:motivation}
\vspace{-0.25in}
\end{figure}

Urban data is unique in that it captures the behavior of the different
components of a city over space and time: its citizens, existing
infrastructure (physical and policies), and the
environment~\cite{brookings}. The availability of these data makes it
possible to not only better understand the individual components but also
obtain insights into how they interact.
When an expert finds an \emph{unexpected} pattern or feature in a
\dataset, other related data may help explain why and under which
conditions the pattern occurs.
Consider the top plot in Figure~\ref{fig:motivation}, which shows the
number of daily taxi trips in New York City (NYC) during 2011 and
2012. While the distribution of trips over time is very similar for the
two years, we observe two large drops: one in August 2011 and another
in October 2012. A natural question is what might have caused these
drastic reductions.  By examining wind speed data (bottom plot in
Figure~\ref{fig:motivation}), we discover that these drops occur on
days with unusually high wind speeds; here, the high wind speeds were 
due to hurricanes Irene and Sandy. This suggests a new
hypothesis to be further investigated: high wind speed leads to
significant reduction in the number of taxi trips.

Besides enabling \emph{hypothesis generation}, studying relationships
among \datasets can also help with \emph{hypothesis testing}.
For instance, the difficulty in finding taxis when it is raining is a
notorious problem in Manhattan. One long-standing hypothesis to
explain this behavior is that taxi drivers set a daily income goal, and
since there is higher demand on rainy days, they reach their goal
faster and stop working earlier. 
Testing for the presence of such a relationship between \datasets---in
this case, NYC taxi data and weather data---can help experts at the
NYC Taxi and Limousine Commission~(TLC) frame appropriate policies to
counter identified problems.

\paragraph{Relationship Discovery}
In this paper, we define and take a first step towards addressing the
problem of discovering potential \textit{relationships} between spatio-temporal \datasets.
We aim to \textbf{\textit{guide}} users in the data analysis and
exploration
process by allowing them to pose \textit{\textbf{relationship queries}}:

\noindent
\begin{boxedminipage}{\linewidth}
\centering
\emph{Find all \datasets related to a given \dataset $\Dspace$.}
\end{boxedminipage}

\noindent To evaluate the above query, we have to first 
determine \emph{when and how two \datasets are related}.
Answering this question in the context of urban \datasets gives rise
to several challenges.
Urban data can be large vertically, containing hundreds of millions to
billions of data points, and horizontally, consisting of several
attributes~\cite{barbosa@bigdata2014}. 
As a point of comparison, five years of taxi data contain 780 million
trips~\cite{tlcdata}, and the
weather \dataset has over 200 attributes~\cite{weatherdata}.
Also, there is a large number of urban \datasets.
NYC alone has published over 1,300 data sets in the
past two years~\cite{nycopendata}, and this is just a small fraction
of the data collected by the city. 
Since a \dataset can be related to zero or more \datasets through
multiple attributes, there is a combinatorially large number of
possible relationships.

This problem is compounded by the fact that these \datasets contain
both spatial and temporal attributes at different resolutions. For
example, values for the weather attributes are collected at hourly
intervals (temporal resolution) for the whole city (spatial
resolution). In contrast, NYC taxi trips are associated with GPS coordinates
with time precision in seconds.
Other \datasets use spatial resolutions at the level of neighborhoods or zip
codes, and temporal resolutions as daily, weekly, and monthly intervals.
Since relationships can materialize at any of these resolutions, they
should be evaluated at multiple resolutions.

The data complexity coupled with the sheer number of available
\datasets and the combinatorially large number of possible
relationships make it hard for domain experts to
comprehend the information and the insights it can potentially offer.
Of the several thousand possible relationships between pairs of
attributes in different \datasets, only a small fraction is actually
informative. Unless known a priori, looking for meaningful
relationships between these \datasets is like, as the clich\'e goes,
``finding a needle in a haystack.''

Another challenge in identifying a possible relationship lies in defining
the conditions implicating such a relationship. For example, consider
the wind speed data and the NYC taxi trip data depicted in
Figure~\ref{fig:motivation}. There is no apparent relation between the
two \datasets during the normal course of time: it is only when the
wind speed is abnormally high (in that case, due to hurricanes)
that we can see a connection with taxi trips.
This is a common pattern observed across urban \datasets,
where relationships become visible only at \textit{spatio-temporal regions}
(locations in space and time) that behave differently compared to
the regions' neighborhood.

Standard techniques, such as
Pearson correlation coefficient~\cite{statistics} or dynamic time
warping~\cite{keogh@2004}, do not capture these relationships
because they ignore the spatio-temporal dependencies inherent
in the data and operate globally over the entire data (see Section~\ref{sec:baseline-comparison}).
Therefore, we need a method that captures the variation of the data
over space and time at different and arbitrary resolutions.

\paragraph{Our Approach}
To address these challenges, we propose the \textbf{\datapolygamy}
framework.
We introduce the notion of \emph{topology-based relationships}, where two
\datasets are related if there is a relationship between the \textit{salient features} of the data.
A salient feature corresponds to a spatio-temporal region that
exhibits an unusual behavior with respect to its neighborhood.
To efficiently identify salient features, we use and extend
techniques from computational topology.
Topology-based techniques are naturally suited for studying properties 
of data involving spatial and geometric domains
(\eg see \cite{DNN13,topoinvis,ZMT05}).
To give some intuition for why and how we apply topology, suppose we model a
time step in an urban data set as a terrain, where the height of each
point of the terrain represents the data value at that spatial
location. In this case, the variation over space is captured by the peaks and
valleys of this terrain. This can be extended to include time by
modeling the data as a high dimensional terrain. The salient features,
which as mentioned earlier correspond to spatio-temporal regions behaving
differently from their neighborhood, are inherently represented
as tall peaks and deep valleys.
Topological methods provide \emph{efficient algorithms to represent
and compute such features}.
In addition, they can identify features that have an \textit{arbitrary
  spatial structure} and \emph{straddle multiple time intervals}; they
are also \textit{generic}, in the sense that they work on data having
different dimensions and resolutions without requiring any
modification.

Given two \datasets, to determine whether they are related, we assess
how similar their corresponding terrains are, i.e., the similarities
in the spatio-temporal variation patterns of the \datasets.
In the \datapolygamy framework, this is accomplished in three steps:

\vspace{-0.1in}
\begin{enumerate}[leftmargin=*,noitemsep]
\item \emph{\DataSet Transformation.} Each attribute of the two
  \datasets is transformed into a \emph{scalar function}.  A scalar
  function provides a mathematical representation of the terrain
  corresponding to a particular attribute of a \dataset.
  
\item \emph{Feature Identification.}  A topological data structure
  is computed for every scalar function, which provides an abstract
  representation of the peaks and valleys of the scalar function. This
  structure is used as an \emph{index to efficiently identify salient features} in the
  data, which are defined based on thresholds that capture the
  extent of normal behavior of the scalar function.
  We develop a new method based on the notion of \emph{topological
  persistence}~\cite{persistencediag} to automatically compute these
  thresholds.  

\item \emph{Relationship Evaluation.}
  Possible relationships are then identified based on feature similarity.
  Our framework filters out relationships that are \textit{not
    statistically significant}.  Since existing Monte Carlo methods
  assume independence across samples, we develop restricted Monte
  Carlo permutation tests that respect data dependencies due
  to spatial and temporal proximity.
\end{enumerate}
  \vspace{-.2cm}
Users can then pose relationship queries over the resulting
relationships. Hypothesis generation is supported by querying for
relationships among all \datasets, while a given hypothesis can be
tested by querying for relationships between the \datasets involved in
the hypothesis.
Sections~\ref{sec:event-relationships}, \ref{sec:event-compute},
and \ref{sec:relationship-query} provide the formal definitions and
describe the algorithms used in these stages. The end-to-end
\datapolygamy framework is presented in Section~\ref{sec:framework}.

Because we consider 
a large number of urban data sets, each containing many attributes,
we need to compute thousands of scalar functions and derive
millions of relationships. However, both salient feature
identification and relationship querying are embarrassingly parallel
operations. 
In
Section~\ref{sec:implementation}, we briefly describe a map-reduce implementation of the
\datapolygamy framework. 

We demonstrate the efficiency and robustness of our framework in
Section~\ref{sec:exp} through an experimental evaluation using over
300 urban \datasets of varying spatio-temporal resolutions.
We also present use cases demonstrating its effectiveness at
identifying informative relationships.

Note that our goal is to support users in the data exploration process
by helping them discover data sets that may be relevant for their
task---similar to a search engine that returns a set of potentially
relevant documents for a given keyword query.
Users can then use the identified relationships for further analysis, such as
testing for spurious relationships, testing for causality (\eg to support a hypothesis), 
and generating new hypotheses.

\paragraph{Contributions}
We define and propose a topology-based approach to the problem of identifying
relationships across a large number of spatio-temporal
data sets. 
Our main contributions are:
\vspace{-0.2cm}
\begin{itemize}[leftmargin=*,noitemsep]
\item We introduce the notion of \emph{topology-based relationships} to
determine whether \datasets are related through salient features.
\item We develop a \emph{scalable framework} that identifies salient features based on
  the topology of the data, and a topology-based index that provides
  an \textit{output-sensitive} strategy to compute these features: the time
  taken is linear in the size of the output. We also propose
  an \emph{algorithm that automatically determines feature thresholds in a
  data-driven fashion}.
\item We define the \emph{relationship operator}, which returns the
  set of statistically significant relationships between two
  \datasets.  To determine whether a relationship is significant, we
  \emph{develop a 
  strategy that applies 
  Monte Carlo
  permutation tests and respects data dependencies due to spatial and
  temporal proximity}.
\item We describe a scalable, map-reduce implementation of the \datapolygamy
  framework.
\item We perform an extensive experimental evaluation, using real and
  synthetic data, which shows that our framework is robust, efficient, and
  effective.
\end{itemize}
\vspace{-0.2cm}

\begin{figure}[t]
\centering
\includegraphics[width=0.8\linewidth]{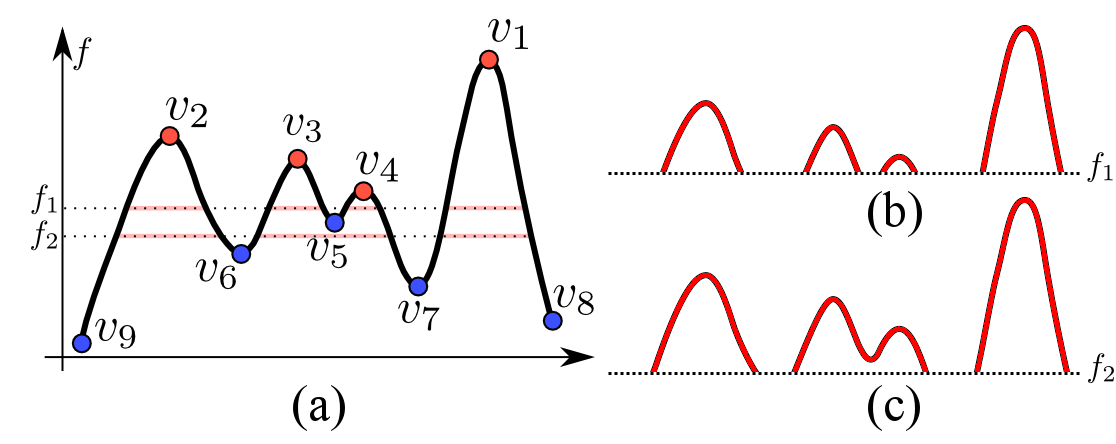}
\vspace{-0.15in}
\caption{
\textbf{(a)}~A sample 1D scalar function. The labeled points form the set of maxima (red) and minima (blue).
\textbf{(b)}~The super-level set at $f_1$ consists of four components.
\textbf{(c)}~The super-level set at $f_2$ consists of three components. 
}
\label{fig:bkgd-1}
\vspace{-0.2in}
\end{figure}

\vspace{-.2cm}
\section{Topology-based Relationships}
\label{sec:event-relationships}

In this section, we provide the required mathematical background and
define the terms used in this paper, which are based on concepts
from computational topology.
We refer the reader to the following
textbooks~\cite{EH09,morsebook} for a comprehensive discussion on
these topics.
We start by  formally defining the concept of a topological feature
and topology-based relationships.
Then, we propose two measures to evaluate
these relationships.

\vspace{-.1cm}
\subsection{Topological Features}
\label{sec:event} 

To discover relationships between two \datasets, we first 
identify the set of topological features
of the \textit{scalar functions} that represent the \datasets.

\paragraph{Scalar Functions}
Let $\Dspace$ be a \dataset, and $A$ an attribute of $\Dspace$.
To identify the set of features with respect to the attribute $A$, we first represent 
the attribute as a \emph{time-varying scalar function}.

\vspace{-.2cm}
\begin{definition}
A \emph{scalar function} $f:\Sspace \rightarrow \Rspace$ maps points
on a spatial domain $\Sspace$ onto a real value.
\end{definition}
\vspace{-.4cm}
\begin{definition}
A \emph{time-varying scalar function} $f:[\Sspace \times \Tspace] \rightarrow \Rspace$ maps points
on a spatial domain across time onto a real value.
\end{definition}
\vspace{-.2cm}

\begin{figure}
\centering
\includegraphics[width=0.9\linewidth]{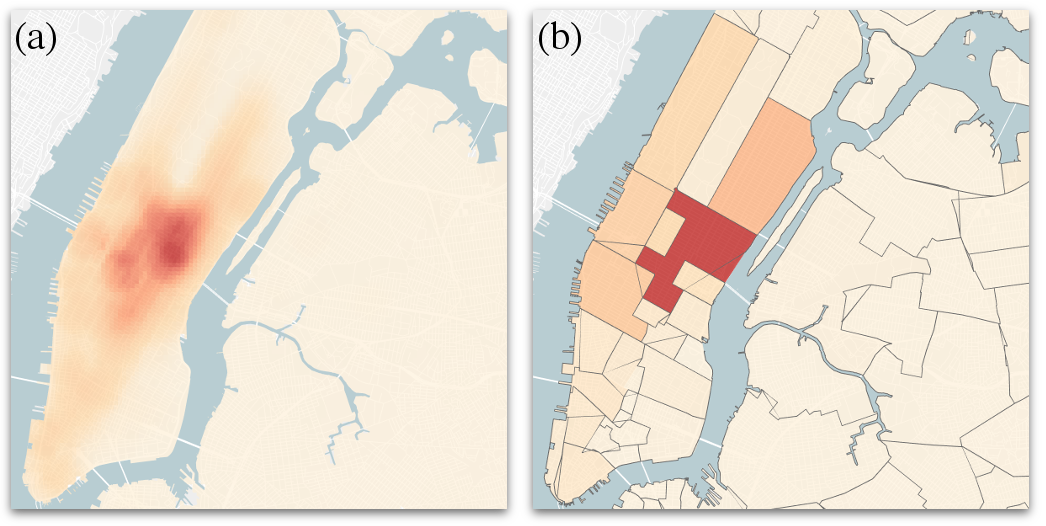}
\vspace{-.1in}
\caption{One time step (2D slice) of the 3D function representing the
  density of taxi trips in NYC at different resolutions. Dark and
  light regions correspond to high and low trip density, respectively.
  \textbf{(a)}~NYC is represented using a high-resolution grid and the
  density is provided for each cell of this grid.  \textbf{(b)}~A
  lower resolution, at the level of neighborhood, is used.  }
\label{fig:taxi-density}
\vspace{-.15in}
\end{figure}

The spatial resolution of the \dataset $\Dspace$ determines the
structure of the spatial domain $\Sspace$.  For example, the NYC
weather \dataset provides information on different climate attributes,
such as temperature, precipitation, and wind speed. The values of
these attributes correspond to an hourly time period for the
entire city, \ie all the values correspond to the same spatial point.
In this case, the domain $\Sspace \times \Tspace$ of the time-varying
scalar function is a simple time series, \ie a 1D function (0D in
space and 1D in time).  Figure~\ref{fig:bkgd-1}(a) and the two time
series shown in Figure~\ref{fig:motivation} are examples of 1D scalar
functions.
On the other hand, the NYC taxi data consists of a set of taxi trips,
each containing the GPS coordinates for pick-up and drop-off
locations.  From these data, we can obtain a distribution of taxi trips
over space and time by partitioning NYC into a set of polygons (\eg
neighborhoods) and counting the trips that start (or end) in each
polygon at different time steps.  This is a \textit{density function},
where the spatial domain is 2D, and thus the time-varying scalar
function is 3D (2D in space and 1D in time).
Figure~\ref{fig:taxi-density} shows the density function at two
different spatial resolutions for one time step (\ie one hour time
period).
The different scalar functions that can be used to represent a
\dataset are discussed in Section~\ref{sec:function}.

Irrespective of the temporal resolution, time always contributes to one dimension in the time-varying scalar function.
Unless otherwise noted, we use the term \emph{scalar function} to
refer to a time-varying scalar function corresponding to a (\dataset,
attribute) pair.

\paragraph{Topological Features}
Interesting features of a scalar function $f$ are captured by the \textit{critical points} of $f$.

\vspace{-.1cm}
\begin{definition}
Given a smooth function $f$, the \emph{critical points} of $f$ are the points where the gradient becomes zero, \ie $\nabla f = 0$. 
\end{definition}
\vspace{-.1cm}

We assume that the scalar function $f$ is a Morse function~\cite{morsebook}. 
A Morse function has the property that (i)~no two critical points have the same function value;
and (ii)~there are no degenerate critical points (\ie $\nabla^2 f \neq 0$).
Any continuous function $f$ can be made Morse via a simulated perturbation of 
$f$ by an infinitesimally small value such that no two points
have the same function value~\cite{Ede01}.  We provide a detailed
discussion on Morse functions in Appendix~\ref{app:morse}.

We are interested in two particular types of critical points, \emph{maximum} and \emph{minimum}, collectively known as \emph{extrema}. Given a Morse function, maximum and minimum points are defined as follows:

\vspace{-.2cm}
\begin{definition}\label{def:cp}
A point $x$ is a \emph{maximum} if $f(x) > f(x')$, $\forall x' \in N(x)$, where $N(x)$ defines the local neighborhood of $x$.
Similarly, $x$ is a \emph{minimum} if  $f(x) < f(x')$, $\forall x' \in N(x)$.
\end{definition}
\vspace{-.2cm}

The red and blue points in Figure~\ref{fig:bkgd-1}(a) correspond to
the set of maxima and minima, respectively.
We use the neighborhood of critical points of a function to represent the topological features of the data. 
The neighborhoods of the maxima and minima of $f$ are captured by the \emph{super-level} and \emph{sub-level sets} of $f$, respectively.

\vspace{-.2cm}
\begin{definition}
Given a scalar function $f$, the \emph{super-level set} at a real value $\theta$ is defined as $f^{-1}([\theta,\infty))$, \ie
the pre-image of the interval $[\theta,\infty)$. The \emph{sub-level set} at $\theta$ is defined as $f^{-1}((-\infty,\theta])$.
\end{definition}
\vspace{-.2cm}

In other words, the super-level set at a real value $\theta$
is the set of all points on the domain of $f$ having function value
greater than or equal to $\theta$.
For example, the super-level set of the function in Figure~\ref{fig:bkgd-1}(a) at function value $f_1$ consists of 4 components (Figure~\ref{fig:bkgd-1}(b)), while the super-level set at $f_2$ consists  of 3 components (Figure~\ref{fig:bkgd-1}(c)). 
Similarly, the sub-level set at $\theta$ is the set of all points in the domain of $f$
having function value less than or equal to $\theta$.

We define two types of features---positive and negative---using super-level and sub-level sets, respectively. 
\vspace{-.1cm}
\begin{definition}
Given a feature threshold $\theta^+$, the set of \textbf{\emph{positive features}} is defined as the super-level set  $f^{-1}([\theta^+,\infty))$.
\end{definition}
\vspace{-.4cm}
\begin{definition}
Given an feature threshold $\theta^-$,  the set of \textbf{\emph{negative features}} is defined as the sub-level set $f^{-1}((-\infty,\theta^-])$. 
\end{definition}
\vspace{-.2cm}

\paragraph{Feature Representation}
The spatial domain $\Sspace$  of $\Dspace$ is represented as a
set of regions $\{s_1,s_2,\ldots,s_n\}$ that partition the spatial
extent of $\Dspace$.
Each region $s_i$ corresponds to a polygon defined by the
resolution used. For instance, the lowest resolution consists of the
space represented as a single region or polygon. By using smaller
polygons to partition the space, one could obtain a higher resolution
representation as shown in Figure~\ref{fig:taxi-density}(a).
The temporal domain $\Tspace$ is represented as a set of time
intervals $\{t, t+\delta, t+2\delta, \ldots \}$.  The temporal
resolution is defined by the value of $\delta$. For example, when
$\delta$ = 1 hour, the function is specified for hourly time steps.

\vspace{-0.2cm}
\begin{definition}
  A \emph{spatio-temporal point} is represented by a (spatial region,
  time interval) pair.
\end{definition}
\vspace{-0.2cm}

Topological features of $f$ correspond to a set of spatio-temporal points over the domain of $f$.
Intuitively, they represent spatio-temporal points where attribute $A$ of 
\dataset $\Dspace$ deviates from its normal behavior, and capture the variation of $A$ over both space and time. 
Here, the thresholds $\theta^+$ and $\theta^-$ define the extent of normal behavior of $A$. 
Salient features of the function can be identified by appropriately setting the values of these thresholds.
For example, using the indicated
values of $\theta^+$ and $\theta^-$ in Figure~\ref{fig:motivation}, features corresponding to the
hurricanes are obtained.
We describe an algorithm to identify
the appropriate values for $\theta^+$ and $\theta^-$ in Section~\ref{sec:persistence}.

\subsection{Feature Relatedness}
\label{sec:relation}

Consider two scalar functions: $f_1(\Dspace_1, A_1)$ corresponding to attribute $A_1$
of \dataset $\Dspace_1$, and $f_2(\Dspace_2, B_1)$ corresponding to attribute $B_1$ of
$\Dspace_2$.
Without loss of generality, we assume that the two functions have the same spatial
and temporal resolution. 
Let $\Sigma_1$ and $\Sigma_2$ be the set of features of $f_1$ and $f_2$, respectively. Let
$\Sigma = \Sigma_1 \bigcap \Sigma_2$ be the set of all spatio-temporal points
that are features in both $f_1$ and $f_2$. The possible relationships between two functions
are defined based on the relationship between their features. 

\vspace{-.2cm}
\begin{definition}
Two functions $f_1$ and $f_2$ are \emph{feature-related} at a spatio-temporal point $x = (s,t)$
if $x \in \Sigma$.
\end{definition}
\vspace{-.2cm}

At points not in $\Sigma$, the two functions are not feature-related.
Let $\Sigma^+_i \subset \Sigma_i$ be a set such that $\forall x \in \Sigma^+_i, x$ is a positive feature. 
Similarly, let $\Sigma^-_i \subset \Sigma_i$ be the set of negative features.

\vspace{-.2cm}
\begin{definition}
$f_1$ and $f_2$ are \emph{positively related} at a spatio-temporal point $x \in \Sigma$
if ($x \in \Sigma^+_1$ and $x \in \Sigma^+_2$) or ($x \in \Sigma^-_1$ and $x \in \Sigma^-_2$).
\end{definition}
\vspace{-.4cm}
\begin{definition}
$f_1$ and $f_2$ are \emph{negatively related} at a spatio-temporal point $x \in \Sigma$
if ($x \in \Sigma^+_1$ and $x \in \Sigma^-_2$) or ($x \in \Sigma^-_1$ and $x \in \Sigma^+_2$).
\end{definition}
\vspace{-.2cm}

For instance, consider the features from Figure~\ref{fig:motivation} corresponding to the indicated thresholds. 
At the spatio-temporal point corresponding to hurricane Sandy, there is a negative feature in the taxi density function 
and a positive feature in the wind speed function. The functions are therefore negatively related at that point.

\subsection{Relationship Score and Strength}
\label{sec:measures}
To assess the nature of the relationship between a given pair of functions,
we define the following two measures.

\paragraph{Relationship Score $\tau$}
We are interested in evaluating 
the overall nature of the relationship between two functions, \ie whether it is
always positive, always negative, or somewhere in between.
To do so, we define the \emph{relationship score} $\tau$ between two functions $f_1~(\Dspace_1,A_1)$ and $f_2~(\Dspace_2,B_1)$.
Let $\Sigma_1$ and $\Sigma_2$ denote the set of features of $f_1$ and $f_2$, respectively. 
As defined earlier, the set $\Sigma = \Sigma_1 \bigcap \Sigma_2$ denotes the feature-relations between the two functions.
Let $\#p$ and $\#n$ be number of positive and negative feature relations
in $\Sigma$. The relationship score is defined as 
\vspace{-.2cm}
\begin{equation}
\tau = \frac{\#p - \#n}{|\Sigma|}
\end{equation}
A value of $\tau$ closer to $+1$ indicates that the two functions are positively related, while a value closer to $-1$ indicates that the functions are negatively related. 

\paragraph{Relationship Strength $\rho$}
This measure is used to capture how frequently features in two
functions are related: the more frequently the features are related,
the stronger the relationship is.  
We model the set of features as binary classifiers.
Consider any spatio-temporal point $x \in \Sigma_1$. If $x$ is also
present in $\Sigma$, it is considered as a true positive~($tp$). A point $x$
is a false positive~($fp$) when $x \in \Sigma_1$ and $x \notin \Sigma$.
Similarly, $x$ is false negative~($fn$) when $x \notin \Sigma_1$ and
$x \in \Sigma$. 
Given these classifiers, we can compute precision and recall as
\begin{eqnarray}
precision &=& \frac{\#tp}{\#tp + \#fp} \\
recall &=& \frac{\#tp}{\#tp + \#fn}
\end{eqnarray} 
Intuitively, $precision$ gives a measure of how often features in $f_1$ are related with features in $f_2$, and
$recall$ gives a measure of how often features in $f_2$ are related with features in $f_1$.
We then use the F1 score to measure the
relationship strength:
\begin{equation}
\rho = F_1(f_1,f_2) = 2 \times \frac{precision \times recall}{precision + recall}
\end{equation}
A value of $\rho$ closer to 1 indicates a strong relationship between the two functions, 
since a feature in one function almost always indicates a feature in the other function as well.
Similarly, a value of $\rho$ close to zero indicates a weak relationship.

\vspace{-.25cm}
\section{Merge Tree Index}
\label{sec:event-compute}

We use a topological data structure called \emph{merge tree} to
efficiently identify salient features corresponding to a scalar
function. In what follows, we give an overview of merge trees and introduce a new algorithm
to compute the feature thresholds $\theta^+$ and $\theta^-$, a
crucial step in this process.

\begin{figure}[t]
\centering
\includegraphics[width=0.8\linewidth]{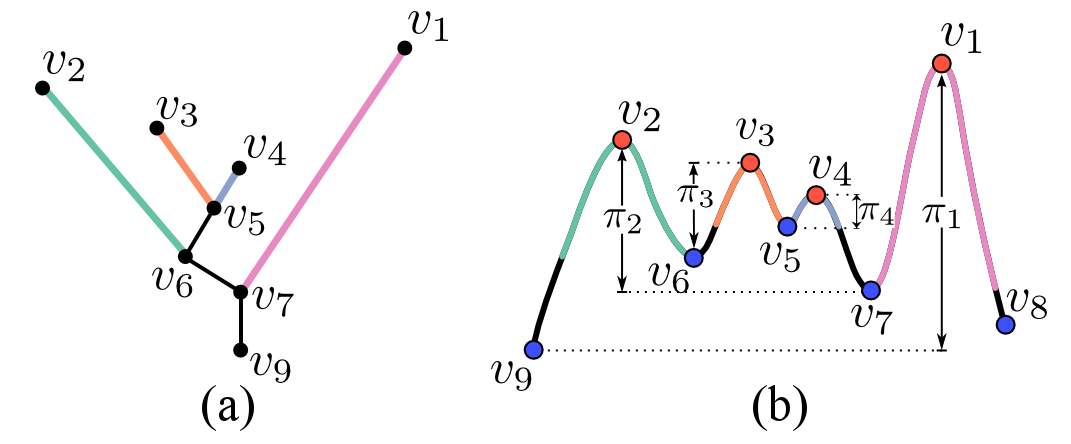}
\vspace{-0.2in}
\caption{
\textbf{(a)}~Join tree of the function shown in Figure~\ref{fig:bkgd-1}.  The edges are colored based on the descending path traversed from the corresponding maxima illustrated in \textbf{(b)}. $\pi_i$ denotes the persistence of maximum $v_i$. 
}
\label{fig:bkgd-2}
\vspace{-0.2in}
\end{figure}

\subsection{Index Creation}
\label{sec:merge-tree}

Recall that a super-level set (or sub-level set) of a scalar function
$f$ at a given function value consists of multiple connected
components.  
Therefore, decreasing or increasing the function values
changes the topology, i.e., the number of connected
components.
A \emph{merge tree} tracks the evolution of super-level sets or
sub-level sets of $f$ with changing function value.  
Formally, there are two types of merge trees.

\vspace{-0.2cm}
\begin{definition}
The \emph{join tree} of $f$ tracks the connected components of the super-level sets of $f$
with decreasing function value.
\end{definition}
\vspace{-0.4cm}
\begin{definition}
The \emph{split tree} of $f$ tracks the connected components of the sub-level sets of $f$
with increasing function value.
\end{definition}
\vspace{-0.2cm}

Consider the 1D function shown in Figure~\ref{fig:bkgd-1}(a). At the highest function value, 
a single super-level set component is created at $v_1$. As we decrease the function value, the number of 
components remain at one until the function value is equal to that of $v_2$. As we keep decreasing the 
function value, two more components are created at $v_3$ followed by $v_4$ (Figure~\ref{fig:bkgd-1}(b)). However, when the function value
reaches $v_5$, the components created at $v_3$ and $v_4$ merge into one component, 
reducing the number of components from 4 to 3 (Figure~\ref{fig:bkgd-1}(c)).
We stop this process when the function value goes below 
$v_9$ (the global minimum). At this point, there is a single super-level set component composed of the entire domain. 
The join tree tracks this evolution as a graph. 
Figure~\ref{fig:bkgd-2}(a) shows the join tree of the 1D function from Figure~\ref{fig:bkgd-1}(a). 
\highlight{
The nodes of the graph correspond to critical points where the number of components change,
while an edge represents the connected super-level set component between its
end points. For example, the edge $(v_2,v_6)$ corresponds to the green connected super-level set component
in Figure~\ref{fig:bkgd-2}(b).}
The root node of a join tree is the global minimum of $f$, while the non-root leaf nodes correspond to the set of maxima
of $f$. 
Similarly, the root node of a split tree corresponds to the global maximum and its non-root leaf nodes correspond
to the set of minima of $f$. 
In order to compute merge trees, we first have to obtain a discrete representation of a scalar function, which we describe next.

\paragraph{Scalar Function Representation}
Consider the spatial domain of a \dataset $\Dspace$ consisting of regions $\{s_1,s_2,\ldots,s_n\}$. 
Let the temporal domain of $\Dspace$ consist of $m$ time steps $\{t_1, t_2,\ldots,t_m\}$.
We create a graph $G=(V,E)$ to represent the spatio-temporal domain of $\Dspace$ as follows.
Vertex $v_{x,z} \in V$ represents the spatio-temporal point corresponding to region $s_x$ at time $t_z$.
Thus, $|V| = n \times m$. The edges $E = E_S \bigcup E_T$ are divided
into two categories: 

\vspace{-0.2cm}
\begin{itemize}[leftmargin=*,noitemsep]
\item \emph{spatial edges}: $ E_S = \{(v_{x,z},v_{y,z})~|~s_x ~\mathrm{adjacent ~to}~ s_y, \forall z \in [1,m]\} \nonumber$
\item \emph{temporal edges}: $E_T = \{(v_{x,z},v_{x,z+1}), \forall x \in [1,n], ~z \in [1,m)\} \nonumber$
\end{itemize}
\vspace{-0.2cm}

\noindent  Edges in $E_S$ connect adjacent regions of the space for each
time step, and edges in $E_T$ connect a region across adjacent
time steps.

We use a piecewise linear (PL) function defined on $G$ to represent
the scalar function $f$: the function is defined on the vertices of
$G$ and linearly interpolated within each edge.
The graph allows a single representation to be used irrespective 
of the dimension of the spatio-temporal domain, thus 
supporting  different resolutions and dimensions of the data.

\paragraph{Merge Tree Computation}
The merge tree of a PL function can be efficiently computed in
$O(N \log N + M\alpha(M))$ time using the union-find data 
structure, where $N$
and $M$ are the number of vertices and edges, respectively, in $G$. 
Since the spatial domains considered in this work correspond to
cities, the graph $G$ representing these domains is planar. Thus,
$M = O(N)$. 
\highlight{
The algorithm to compute join trees is given in
Procedure~\ref{code:join-tree} (for more details, see Appendix~\ref{app:merge-tree}).
The split tree is computed analogously by using the function $f' = -f$ in this algorithm.
}

\subsection{Querying Features}
\label{sec:event-querying}

We use the join and split trees as indices to efficiently compute the
set of features, \ie the super-level sets and sub-level sets,
respectively.
Let $\theta$ be the feature threshold.  The algorithm to compute the
super-level set $f^{-1}([\theta, \infty))$ using the join tree $J_T$
is as follows: \vspace{-0.05in}
\begin{enumerate}[leftmargin=*,noitemsep]
\item Identify the set $V^+ = \{v~|~v$ is a maximum and $f(v) \geq
  \theta\}$. 
This is accomplished by going over the non-root leaf nodes of $J_T$.
\item Set $\Sigma^+ = \emptyset$.
\item While $V^+ \neq \emptyset$
\vspace{-0.1cm}
\begin{enumerate}[noitemsep]
\item Remove $v$ from $V^+$ and add to $\Sigma^+$.
\item Let $L^- = \{u | f(u) \leq f(v)$ and $u$ is adjacent to $ v\}$.
\item Add $u \in L^-$ to $V^+$ if $\theta \leq f(u)$.
\end{enumerate}
\item The set $\Sigma^+$ contains the vertices of $G$ that belong to the super-level set at $\theta$.
\end{enumerate}
\vspace{-0.05in}
The algorithm 
performs a descending path traversal of adjacent vertices from the set of valid 
maxima (having function value greater than $\theta$) until the required threshold is reached.
The colored regions in Figure~\ref{fig:bkgd-2}(b) indicate the descending paths followed by the algorithm
starting from the different maxima of the function shown in Figure~\ref{fig:bkgd-1}(a). 
This is analogous to traversing down the edges of the join tree.
The sub-level set at $\theta$ is computed similarly,
through an ascending path traversal starting from the minima of the
function.

\paragraph{Time Complexity}
Since the vertices of $G$ are sorted when computing the join (or split) tree, the critical points of the function
are also stored in sorted order. Thus, the number of comparisons required to identify 
$V^+$ (or $V^-$) is $|V^+|$ (or $|V^-|$). Each descending (or ascending) path traversal stops as soon as it reaches a vertex $u$ that is not a feature.
Thus, the number of  vertices touched during querying is $O(\Sigma^+)$ (or $O(\Sigma^-$)). In other words,
given the join and split trees, feature identification for a given threshold is \textbf{\textit{output-sensitive}}.

\subsection{Feature Threshold Computation}
\label{sec:persistence}

Intuitively, our goal is to classify topological features not adhering
to normal behavior as \emph{salient features}.  We are also interested
in identifying \emph{extreme features}, which correspond to outliers
among salient features. For instance, the extremely high wind speeds
during a hurricane correspond to extreme features

While users with domain knowledge can provide thresholds for computing
features, this might not be feasible over all \datasets. Thus,
we devise a data-driven approach to identify the required thresholds,
$\theta^+$ and $\theta^-$.
Our approach is inspired by the
\emph{persistence diagram}~\cite{persistencediag}, which is commonly
used in scientific visualization applications to visually identify
meaningful thresholds~\cite{DNN13}.  However, instead of relying on
users to visually select thresholds, we develop an algorithm to
automatically identify them.

\setlength{\textfloatsep}{0pt}
\begin{procedure}[t]
\footnotesize
\highlight{
\caption{ComputeJoinTree()}
\label{code:join-tree}
\begin{algorithmic}[1]
\REQUIRE Graph $G(V,E)$, Function $f$
\STATE Sort $V$ in descending order of $f$
\FOR{each $v \in V$}
	\STATE $L^+ = \{u | (v,u) \in E$ and $f(v)<f(u)\}$
	\STATE $C$ = $\{$Component$(u) | u \in L^+\}$
	\IF[$v$ is a maximum and creator]{$|C| = 0$} 
		\STATE Create a new join component $C_J$
		\STATE Set Head($C_J$) = $v$, Creator($C_J$) = $v$
	\ELSIF[$v$ is not critical]{$|C| = 1$}
		\STATE Add $v$ to $C$
	\ELSE[$v$ is a destroyer, $|C| = 2$ for Morse functions]
		\STATE Let $C = \{C_1, C_2\}$, $f($Creator$(C_1)) < f($Creator$(C_2))$
		\STATE Merge Components $C_J = C_1 \bigcup C_2$
		\STATE Let $u_1 =~$Head$(C_1)$, $u_2 =~$Head$(C_2)$
		\STATE Add edges $(u_1,v)$ and $(u_2,v)$ to Join Tree $J_T$
		\STATE Set Creator($C_J$) = Creator($C_1$), Head($C_J$) = $v$
		\STATE \label{line:pairing}Pair Creator($C_2$) with destroyer $v$
	\ENDIF
\ENDFOR
\RETURN Join Tree $J_T$
\end{algorithmic}
}
\end{procedure}

\paragraph{Topological Persistence}
Consider a sweep of the function $f$ in decreasing order of function value. As mentioned earlier, the topology of the super-level set changes at critical points during this sweep. 
In particular, at a critical point, either a new super-level set component is created (maximum) or an existing super-level set component is destroyed.
A critical point is a \textit{creator} if a new component is created, and a \textit{destroyer} otherwise. 
Again, consider the example in Figure~\ref{fig:bkgd-2}. At critical point $v_5$, the components created at $v_3$ and $v_4$ are merged into one. In this case, the component created last, at $v_4$, is considered to be destroyed at $v_5$. 
Similarly, one can pair up each creator $c_i$ uniquely with a
destroyer $d_i$ that destroys the topology created at $c_i$.
\highlight{
Note that this pairing can be accomplished while computing the merge tree itself (Line~\ref{line:pairing} in Procedure~\ref{code:join-tree}).
}
The \emph{persistence value} of $c_i$ and $d_i$ is defined as
$|f(d_i) - f(c_i)|$, which indicates the lifetime of the feature
created at $c_i$. Figure~\ref{fig:bkgd-2}(b) illustrates the
persistence values of the different critical points (as $\pi_i$)
together with the creator-destroyer pairs.
Intuitively, the persistence of a maximum (minimum) is equal to the height (depth) of the corresponding peak (valley).

\begin{figure}[t]
\centering
\includegraphics[width=\linewidth]{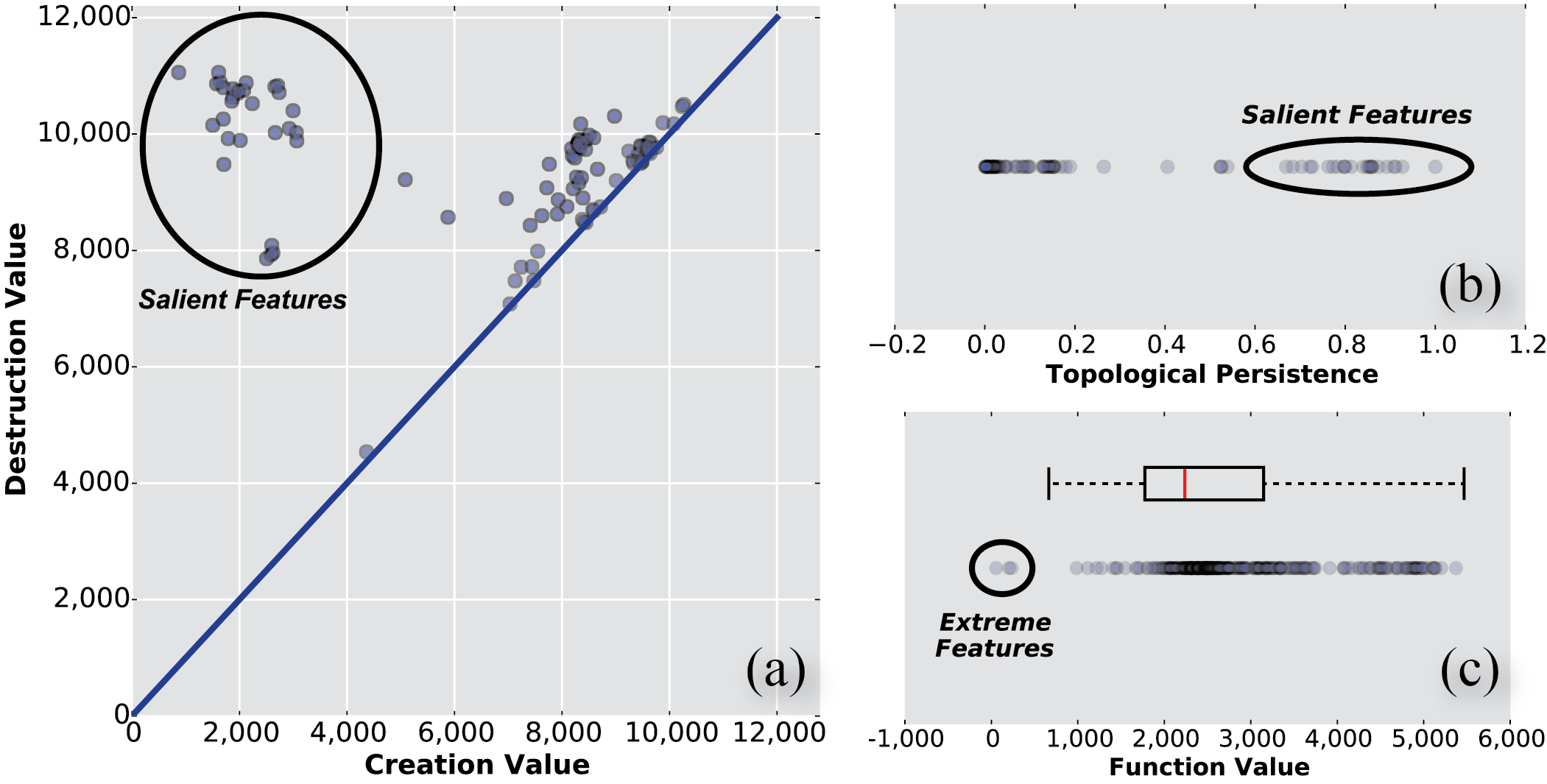}
\vspace{-0.25in}
\caption{
\textbf{(a)}~The persistence of a minima in a persistence diagram is the height above the $x = y$ line.
\textbf{(b)}~A scatter plot 
of the persistence of the minima.
\textbf{(c)}~When considering only negative features across all time intervals, note that function values corresponding to extreme features (\eg during hurricanes) are outliers of the distribution.}
\label{fig:pdiag}
\vspace{-0.20in}
\end{figure}

\paragraph{Thresholds for Salient Features}
The \emph{persistence diagram}~\cite{persistencediag,ELZ02} plots the maxima (or minima) of the input function as a set of 
points on a 2D plane, where the $x$ and $y$ coordinates of a feature correspond to its creation and 
destruction value, respectively. 
Figure~\ref{fig:pdiag}(a) shows the persistence diagram of the minima of the taxi-density function
from Figure~\ref{fig:motivation} corresponding to one month in the data.
Note that the minima are clearly split into \emph{two groups}: those with high persistence (enclosed by the circle) and 
those with low persistence. This split becomes even more prominent when we plot just the persistence values~(Figure~\ref{fig:pdiag}(b)).
A minimum (maximum) with a higher persistence value is considered important since the 
sub- (or super-) level set component created at that extremum has a longer lifetime.
Our goal is to select a threshold $\theta^-$ such that all the high
persistence minima are identified as salient. To do this
automatically, we perform $k$-means clustering with $k = 2$ and use
the highest value over all minima in the high persistence cluster as
the threshold $\theta^-$.  This ensures that all the high-persistence
minima will have function value less than or equal to $\theta^-$, and
will hence be identified as salient features.
$\theta^+$ is identified in a similar manner using the persistence of the set of maxima.

\paragraph{Thresholds for Extreme Features}
Typically, a minimum (maximum) corresponding to an extreme feature will have
function value that is significantly smaller (larger) than those
corresponding to salient features. For example, in
Figure~\ref{fig:pdiag}(c), the function value of minima corresponding
to the extremely low number of taxi trips in NYC between 2009 and 2013
is significantly different from the function value of other minima
(corresponding to salient features).
In order to identify the appropriate thresholds, we first compute the
minima (or maxima) across all time steps that correspond to salient
features. Next, we identify the outlier threshold from this
distribution. We use the standard box
plot thresholds, \ie $Q_1 - 1.5 \times IQR$ for minima
($Q_3 + 1.5 \times IQR$ for maxima), as the required thresholds, where
$Q_1$ and $Q_3$ are the first and third quartile, and $IQR$ is the
inter-quartile range.  The box plot (and the corresponding outlier
threshold) for the extreme negative features corresponding to the taxi
density function is illustrated in Figure~\ref{fig:pdiag}(c).

\paragraph{Adjusting for Seasonal Variations}
Processes in a city are typically dependent on the time of year. For
example, zero depth of snow during summer is normal, while this could indicate an important phenomena during the winter. Thus, it is
important to take into account seasonal variations when computing
features. 
Depending on the temporal resolution, the time range of
a data set is divided into smaller intervals, and the threshold for a
given interval is computed based on the persistence of the extrema present
in that interval. 
For example, we could use monthly or quarter-yearly intervals.

\vspace{-.3cm}
\section{Relationship Operator}
\label{sec:relationship-query}

In the previous sections, we discussed how relationships between functions are identified and measured. We now define the relationship operator, \textsc{relation}$(\Dspace_1,\Dspace_2)$, used to compute the relationship between \datasets $\Dspace_1$ and $\Dspace_2$.
Let $\Dspace_1$ be represented by $n$ functions $\{f_1,f_2,\ldots,f_n\}$, and $\Dspace_2$ by
$m$ functions $\{g_1,g_2,\ldots,g_m\}$. There are $n \times m$ possible relationships
between the two \datasets. 
Since many of these relationships could be due to random chance,
the relationship operator
returns the set of statistically significant relationship pairs $(f_i,g_j)$ together with their corresponding relationship score and strength.

To assess the statistical significance of a potential relationship pair $(f_i,g_j)$, we design Generalized Monte Carlo significance tests \cite{besag1989generalized, ernst2004permutation, fortin2000randomization}.
Let $\Sigma_{1}$ and $\Sigma_{2}$ be the features corresponding to $f_i$ and $g_j$, respectively. 
The null and alternative hypothesis 
are:
\vspace{-0.15cm}
\begin{description}[leftmargin=*,noitemsep]
\item[$H_0$:] The two functions $f_i,~g_j$ are independent in their features $\Sigma_{1}, \Sigma_{2}$.
\item[$H_1$:] The two functions are dependent in their features.
\end{description}
\vspace{-0.15cm}
We examine if we can reject the null hypothesis and accept $H_1$ for any pair of functions based on the identified features and their corresponding relationship score. 
The $p$-value from the Monte Carlo randomization test with test statistic $x^*$ is given by:
\vspace{-0.2cm}
\begin{equation}
p = P(X\leq x^*|H_0) = \frac{\sum_i^{{N}}\mathrm{I}(x_i \leq x^*)}{{N}}~~\mathrm{as}~~N \rightarrow \infty
\end{equation}
where $\mathrm{I}(\cdot)$ is the indicator function and $N$ the number of permutations on the input. 
Given a significance level $\alpha$, the $p$-value is then used to define a statistically significant relationship as follows:
\vspace{-0.2cm}
\begin{definition}
The relationship between two functions $f_i$ and $g_j$ is statistically significant if $p \leq \alpha$.
\end{definition}
\vspace{-0.2cm} 

\setlength{\textfloatsep}{\oldtextfloatsep}

\highlight{Urban \datasets have spatial and temporal dependencies (e.g., due to neighborhood and seasonal effects) that need to be accounted for when designing randomization tests. It is well-known in the statistics literature that, if we ignore these dependencies, a simple Monte Carlo procedure for assessing statistical significance would fail and lead to erroneous claims~\cite{besag1989generalized,manly1997randomization}. To account for the spatio-temporal correlation, a plethora of Monte Carlo and Bootstrap techniques have been developed over the last decades ranging from the block-bootstrap \cite{kunsch1989jackknife} to general restricted Monte Carlo techniques \cite{fortin2000randomization,manly1997randomization} such as the one we propose in this paper.}

\paragraph{Restricted Monte Carlo Tests for Spatial Correlation}
We develop restricted permutation tests that respect the degree of spatial correlation of our data sources. This is typically achieved by designing toroidal shifts, where a function $f$ is wrapped around a two-dimensional torus by connecting the margins, or spatial extents, of the data. Then, a linear map $m$---that maps the torus onto a rotation of itself---will yield a new randomization that still respects any horizontal interactions~\cite{fortin2000randomization,manly1997randomization}. 

However, given the irregular structure of a city, which is an arbitrary non-convex polygon, wrapping the spatial region over a torus is not straightforward. If we consider the spatial domain as a graph, a toroidal shift basically ensures that the adjacency of the non-boundary vertices are maintained. We make use of this observation to devise a toroidal shifting strategy that is applicable to arbitrary graphs. Given a graph $G$ representing the spatial domain, we define the map $m_i: G \rightarrow G$ as follows. We start with a random mapping $m_i(u) = v$. The adjacent vertices of $u$ are then assigned the vertices adjacent to $v$ where applicable. This process is repeated in a breadth-first fashion. This process ensures that, in most cases, the distance between two vertices in $G$ is the same as the distance between them in $m_i(G)$.
Using the above mapping, the restricted Monte Carlo test now becomes:
\begin{equation}
\label{MCtest}
p = \frac{\sum_k^{{|m|}}\mathrm{I}({\tau(f_i, g_j)}_k \leq {\tau(f_i, g_j)}^*)}{{|m|}}
\end{equation}
where ${\tau(f_i, g_j)}_k$ is the relationship score between the two functions $f_i, g_j$ in toroidal shift $k$, and $|m|$ is the total number of toroidal shifts, which affects the power of the statistical test (we use $|m| = 1,000$).

\paragraph{Restricted Monte Carlo Tests for Temporal Correlation}
For 1D functions that are purely temporal and have no spatial domain, 
we wrap time to a one-dimensional torus while rotating the resulting circle to obtain randomizations that respect temporal correlations~\cite{fortin2000randomization}. We then proceed similarly to Equation~\ref{MCtest}.
Unless otherwise mentioned, a relationship implies a statistically significant relationship for the remainder of the paper.

\section{Data Polygamy Framework}
\label{sec:framework}

In this section, we describe the data polygamy framework.
We start by presenting the \emph{scalar functions} that are derived
from a given \dataset and how the framework handles different
resolutions.
Then, we discuss how the data is indexed and queries are evaluated.
Finally, we briefly describe a map-reduce implementation of the framework.

\subsection{Types of Scalar Function}
\label{sec:function}

Consider a \dataset $\Dspace$ having attributes $\{K, S, T, A_1, A_2, \ldots, A_k\}$. 
Let $K$ be an optional unique identifier (the \dataset can have zero or more identifier attributes);
$S$ and $T$ be the spatial and temporal attributes, respectively;
and $A_i, 1 \leq i \leq k$ be numerical attributes.
We are interested in identifying scalar functions that 
capture the different properties corresponding
to a given \dataset. 
For instance, when considering the taxi data, the number of taxis in different locations over time captures the
activity of the taxis, while the attribute corresponding to the fare
captures 
fare patterns
over time and space.
We therefore derive two types of scalar functions to represent
$\Dspace$: \emph{count functions} and \emph{attribute functions}.
Extensions to other types of scalar functions are discussed in Section~\ref{sec:conclusion}.

\paragraph{Count Functions}
\emph{Count functions} are used to capture the activity of the entity represented by the \dataset. 
More formally, consider a spatio-temporal point $(s,t)$. Let $\Gamma$ be the set of tuples in $\Dspace$
such that $S(r) = s$ and $T(r) = t$, $\forall r \in \Gamma$. Here, $S$ and $T$ 
represent the spatial and temporal attributes of a tuple $r$.
We define two types of count functions: \emph{density} and \emph{unique}. The \emph{density function} assigns the value $|\Gamma|$ to the
spatio-temporal point $(s,t)$. For example, the density function of the taxi data assigns the number of trips originating 
at $s$ during the time period $t$ to the point $(s,t)$.
The \emph{unique function} assigns a value equal to the number of unique identifiers of $\Gamma$ to the spatio-temporal
point. For instance, each tuple in the taxi data consists of an identifier corresponding to the medallion of the taxi. 
Thus, the number of unique medallions in $\Gamma$ is essentially the number of unique taxis 
that are present at $s$ during time $t$.
Note that there is one unique function corresponding to each identifier attribute of the \dataset.

\paragraph{Attribute Functions}
For a given attribute $A$, the \emph{attribute function} assigns the
average value of $A(r)$ over all tuples $r \in \Gamma$ to the
corresponding spatio-temporal point $(s,t)$; the function represents
the variation in the properties of a given attribute over space and
time.

\begin{figure}[t]
\centering
\includegraphics[width=0.95\linewidth]{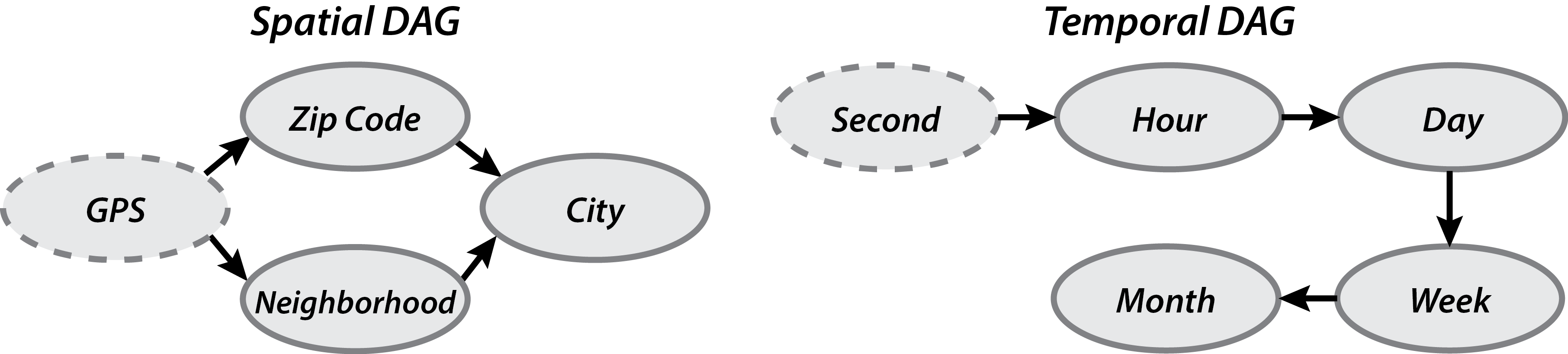}
\vspace{-0.1in}
\caption{Hierarchical relationship for spatio-temporal resolutions
represented by DAGs. Resolutions depicted using solid lines are used for evaluating relationships.}
\label{fig:dag}
\vspace{-0.2in}
\end{figure}

\paragraph{Handling Different Data Resolutions}
It is important that our framework identifies relationships that occur at different
resolutions. As illustrated in Figure~\ref{fig:dag}, these resolutions
are represented as a directed acyclic graph (DAG), where the edges are
directed from a higher resolution to a compatible lower resolution.
The compatibility indicates the ability to convert the data from a
higher resolution to a coarser resolution.
For example, GPS resolution can be transformed into all of the other
resolutions. On the other hand, neighborhood and zip-code resolutions,
being incompatible, can be converted only into the city resolution.
To evaluate the relationship between two functions having different resolutions, 
we first transform both functions into the same compatible resolution, and then evaluate the two functions at this resolution.

\subsection{Indexing and Feature Identification}
\label{sec:indexing-comp}
Given a \dataset $\Dspace$, we first compute all possible 
scalar functions (\ie count and attribute functions) of $\Dspace$ that cover
every viable spatio-temporal resolution. 
For example, if $\Dspace$ is available at a spatial resolution of GPS locations
and temporal resolution of second, then each attribute can be aggregated into 3
spatial resolutions (\ie zip code, neighborhood, and city) and 4
temporal resolutions (\ie hour, day, week, and month), thus resulting in 
a total of 12 spatio-temporal resolutions for which the scalar functions are computed. 
The merge tree index is then built for each 
scalar function. This ensures that all resolutions are considered when
executing a relationship query.
Recall that the computation of feature thresholds takes seasonal
variations into account (Section~\ref{sec:persistence}). In
particular, we use monthly and quarter-yearly intervals when the
temporal resolution is hourly and daily, respectively.
Since thresholds are fixed for a given function, 
to speed up query evaluation,
we pre-compute
and store the features (salient and extreme).

\begin{table*}[t]
\caption{Properties of the \datasets in the \urbannyc collection.}
\vspace{0.1cm}
\centering
\fontsize{7}{7.7}\selectfont
\begin{tabular}{ l r r r r r r p{5.4cm} }
    \hline
    \header{\DataSet} & \header{Data\\Size} & \header{\# Records} & \header{Time Range} & \header{\# Scalar\\Functions} & \headergap{Spatial\\Resolution} & \headergap{Temporal\\Resolution} & \header{Description} \\
    \hline
    Gas Prices & 13 KB & 749 & 2000--2014 & 2 & City & Week & Average
                                                              gasoline
                                                              price in
                                                              dollars
                                                              per
                                                              gallon
                                                              for
                                                              NYC\\ \hline
    Vehicle Collisions & 47 MB & 330 K & 2012--2014 & 11 & GPS &
                                                                 Second & Traffic collision data provided by NYPD\\ \hline
    311 Complaints & 574 MB & 7.40 M & 2003--2014 & 1 & GPS & Second &
                                                                       Records from 311, a telephone number that provides non-emergency services to the city\\ \hline
    911 Calls & 2.2 GB & 6.75 M & 2012--2013 & 1 & GPS & Second &
                                                                  Records from 911, a telephone number that provides emergency services to the city\\ \hline
    Citi Bike Data & 1.6 GB & 10.40 M & 2013--2014 & 5 & GPS
                                                                                                                                                        & Second & Trip data from NYC's bike sharing system\\ \hline
    NCEI Weather Data & 304 MB & 64 K & 2010--2014 & 228 & City & Hour
                                                                                                                                                                                           & Comprehensive weather data from NCEI\\ \hline
    Traffic Speed & 17 GB & 395 M & 2009--2012 & 2 & GPS & Hour &
                                                                  Average speed in the streets of Manhattan\\ \hline
    Taxi Data & 108 GB & 868 M & 2009--2013 & 13 & GPS &
                                                                   Second & Trip data from taxicabs provided by NYC TLC\\ \hline
    Twitter & 656 GB & 1.10 B & 2012--2014 & 5 & GPS & Second & Data obtained from Twitter's public streams\\ \hline
\end{tabular}
\vspace{-0.3in}
\label{tab:nycurban}
\end{table*}

\subsection{Query Evaluation}
\label{sec:query-comp}

Let $\mathcal{D} = \{\Dspace_1, ..., \Dspace_n\}$ be the corpus 
of \datasets that have been indexed. 
We support the general form of the \textit{relationship query}:

\noindent
\begin{boxedminipage}{\linewidth}
\centering
\emph{Find relationships between $\mathcal{D}_1$ and $\mathcal{D}_2$ satisfying \textsc{clause}}
\end{boxedminipage}

\noindent In this query, $\mathcal{D}_1$ and $\mathcal{D}_2$ are collections of \datasets
such that $\mathcal{D}_1 \subseteq \mathcal{D}$ and $\mathcal{D}_2 \subseteq \mathcal{D}$.
If $\mathcal{D}_2 = \varnothing$,
it is assumed that $\mathcal{D}_2 = \mathcal{D}$.
When a relationship query is issued, the \textsc{relation} operator is
applied to all pairs $(\Dspace_i,\Dspace_j)$ of \datasets, such that
$\Dspace_i \in \mathcal{D}_1$, $\Dspace_j \in \mathcal{D}_2$, and
$\Dspace_i \neq \Dspace_j$.  The operator uses the pre-computed set of features to
assess the relationship between the \datasets.
Note that, when considering a pair of functions, the relationship
between them is evaluated for all possible resolutions starting with
the \emph{highest common} resolution.
For example, if the spatial resolutions of two functions are neighborhood and zip code,
then their relationship is evaluated at the city scale 
for different possible temporal resolutions. 
This evaluation is performed for both salient and extreme features.

Computing relationships at different resolutions is important as
scalar functions may relate differently depending on how they are
aggregated.  For example, an hourly resolution might capture
variations within a day, but could miss significant variations across
different days, which can be captured using a daily resolution (see
Section~\ref{sec:qual-evaluation} for an example).

The query returns related scalar function pairs that are statistically significant,
together with the resolutions for which the relationships hold.
The significance level $\alpha$ is set at the commonly used value of 5\%~\cite{ernst2004permutation}.
In the \textsc{clause} for a query, optional condition parameters can
be specified to filter relationships satisfying a minimum score $\tau$
and/or strength $\rho$.
Feature thresholds for computing salient and extreme features can also be optionally 
specified as part of the \textsc{clause} if the user is familiar with any of the \datasets.
When these thresholds are specified,
features are first identified using the merge tree index before evaluating the relationship.

\begin{figure}[t]
\centering
\subfigure[City]{\label{fig:index_1d}
  \includegraphics[height=0.13\textheight]{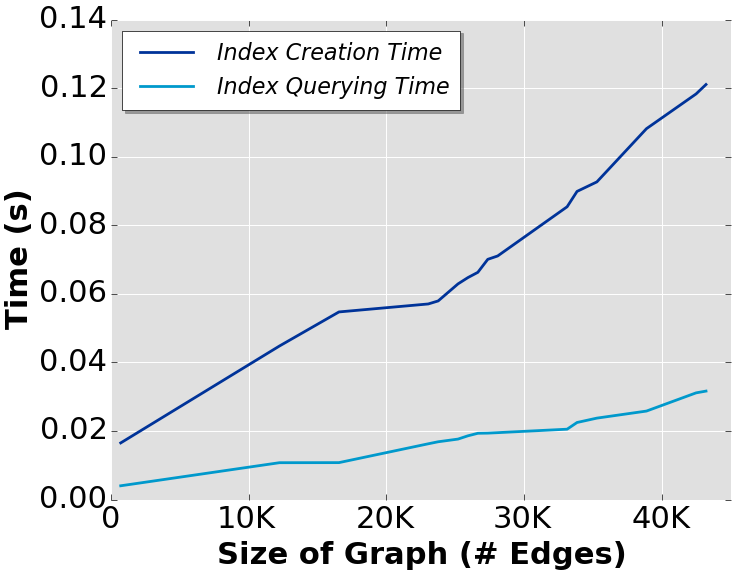}}~
\subfigure[Neighborhood]{\label{fig:index_2d}
  \includegraphics[height=0.13\textheight]{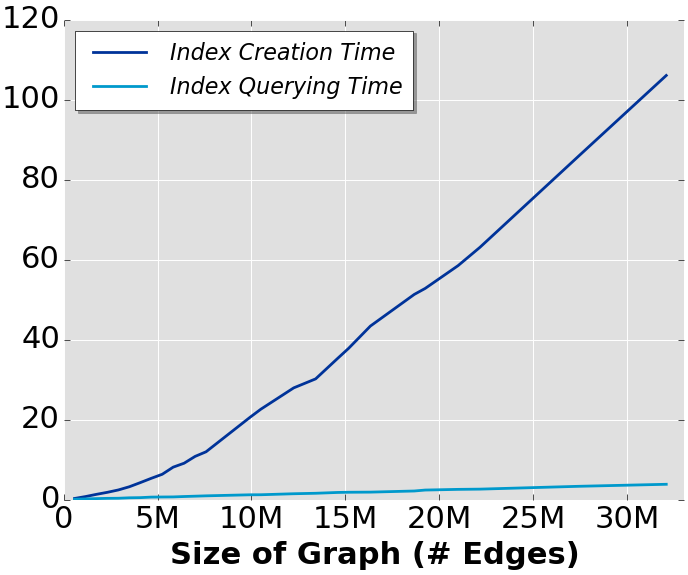}}
\vspace{-0.15in}
\caption{Merge tree index creation and feature querying time.
}
\label{fig:index}
\vspace{-.2in}
\end{figure}

\subsection{Implementation}
\label{sec:implementation}

Given a large number of data sets, such as the urban data sets with which we
experimented (Section~\ref{sec:exp}), thousands of scalar
functions have to be computed, and the number of relationships to be
evaluated during querying is 
in the order of millions.
However, the indexing and querying operations can be run independently for each
scalar function and scalar function pair. 
To leverage the \emph{embarrassingly parallel} nature of these computations, 
we implemented the framework using \mapreduce.
We use three \mapreduce jobs: 
(1)~\textit{Scalar Function Computation} generates all possible scalar functions at different resolutions;
(2)~\textit{Feature Identification} creates the merge tree indexes and identifies the set of features of the different functions; and
(3)~\textit{Relationship Computation} evaluates the relationships between the pairs of functions corresponding to a given query.
Implementation details can be found in Appendix~\ref{app:implementation} and 
the released code.\footnote{\texttt{https://github.com/ViDA-NYU/data-polygamy}}

\paragraph{Space Requirements}
The space required to store a scalar function for a given resolution is equal to the number of vertices in the
graph $G$ representing the domain. For a resolution at the city level
and hourly intervals, this corresponds to the number of time steps, which
is approximately 35~KB ($365 \times 24$ float values) per
year.
For higher spatial resolutions, the space required to store the
function is approximately $n \times 35$~KB, where $n$ is the number of
polygons used to partition the domain. For example, in NYC, for zip
code and neighborhood resolutions, $n \approx 300$.
Typically, the space needed to store scalar functions over all
resolutions is significantly less than the original data itself. As a
point of reference, the 5 years of the raw taxi data takes up 108~GB of space.
In contrast, the 13 possible scalar functions over 8 resolutions uses
only 417~MB.
The size of the merge tree index is proportional to the number of
critical points of the function. While the number of critical points
is bounded by the size of the input graph in the worst case, in
practice it is significantly smaller.
Similarly, even the number of features, while being bounded by the size
of the input graph, is usually much smaller.  For example, storing all
features (salient and extreme) for the taxi data over all different
resolutions takes only 8~MB.

\section{Experimental Evaluation}
\label{sec:exp}

We have performed an extensive evaluation to assess different aspects
of the \datapolygamy framework.
We carried out a controlled experiment to quantitatively evaluate the
correctness and robustness of the relationship operator, and we also
used real-world \datasets to study efficiency and effectiveness
characteristics.
Efficiency was measured to show the feasibility of computing
relationships over a large number of data sets. Effectiveness was
evaluated in two different ways: to demonstrate that the framework is
able to prune spurious relationships, thus reducing the exploration
space presented to users, and to show that our approach uncovers
\emph{interesting}, non-trivial relationships.
\highlight{We also analyzed the relationships obtained from standard correlation techniques
and discuss their shortcomings in identifying interesting relationships.}

\paragraph{Experimental Setup} We used Apache
Hadoop~2.2.0 and Java 1.7.0.  The \mapreduce jobs were executed on a
cluster with 20 compute nodes, each node having an AMD Opteron(TM)
Processor 6272 (4x16 cores) running at 2.1GHz, and 256GB of RAM.

\begin{figure}[t!]
\centering
\subfigure[\urbannyc]{\label{fig:nyc_index}
  \includegraphics[height=0.13\textheight]{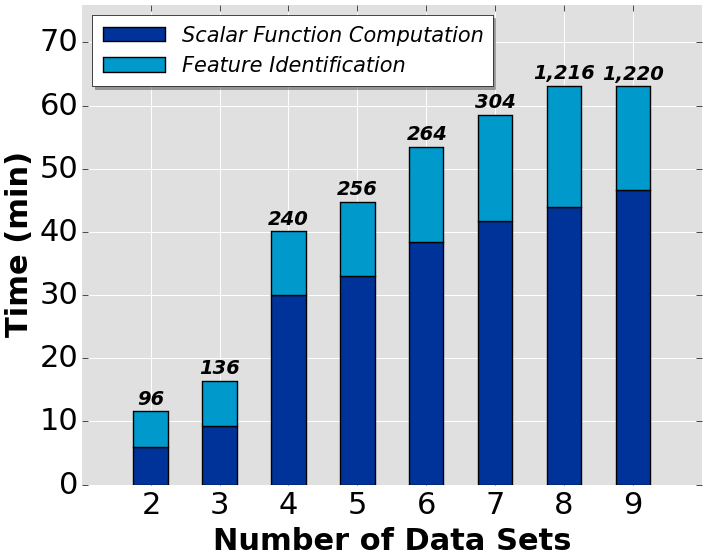}}~
\subfigure[\opendata]{\label{fig:open_index}
  \includegraphics[height=0.13\textheight]{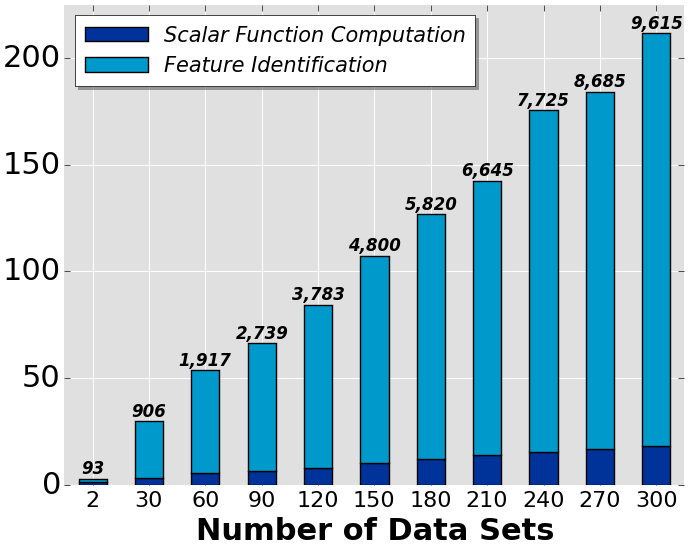}}
\vspace{-0.15in}
\caption{Performance of feature indexing and identification.}
\label{fig:indexcreation}
\vspace{-.3in}
\end{figure}

\paragraph{\DataSets} 
We used two collections of \datasets in our experiments.  The
\urbannyc collection consists of nine urban \datasets obtained from
different NYC agencies or gathered through publicly-available APIs.
These \datasets have been used by various domain experts (mostly in
isolation) for different analyses (see,
e.g.,~\cite{Wood2015,taxivis@tvcg2013})
and are thus useful to evaluate the effectiveness of our
framework at identifying meaningful relationships.
Table~\ref{tab:nycurban} describes these \datasets and their
properties. They vary in size from a few KBs to hundreds of GBs and
have different temporal and spatial resolutions.

The second collection, referred to as \opendata, was primarily used to
test the performance of our framework. 
It consists of 300 spatio-temporal \datasets from NYC Open
Data~\cite{nycopendata}.  Even though most of these \datasets are
relatively small in size (less than 1 GB), the sheer number of
\datasets and the number of attributes they contain (on average, 8 attributes per
\dataset) results in over 2.4 million possible
relationships for a single resolution. 

Each data set in these collections consists of a set of tuples having
metadata about the spatial, temporal, numerical, and identifier attributes.
We use this metadata to perform an additional pre-processing step that
selects data corresponding to these attributes and feeds it to the
scalar function computation module.

\vspace{-.2cm}
\subsection{Performance Evaluation}
\label{sec:quant-evaluation}
\vspace{-.1cm}
 
\paragraph{Indexing and Feature Identification}
To assess the efficiency of feature identification and indexing, we studied
the performance of the merge tree index for a single data set as well
its behavior for an increasing number of data sets.
Figure~\ref{fig:index} plots the running times to create the index
and query for features for the Taxi data (using its density function),
for both city (1D) and neighborhood (3D) resolutions. Here, we used a single node in the cluster.
The plots indicate that the time for creating the merge tree and
identifying features is almost linear in the size of the function (\ie
number of edges in the spatial domain graph).  Note that the indexing
time includes the creation of both join and split trees,
and the querying time includes the computation of thresholds 
as well as the identification of negative and
positive features.
Even for an input having more than 30 million edges, the operations
took less than 2 minutes.
This shows that our approach is scalable and able to handle large data
sets.

The indexing component also performs well as the number of \datasets increases.
This is shown in Figure~\ref{fig:indexcreation}. 
The numbers on the bars indicate the total number of computations
performed. Recall that scalar functions are computed for all
attributes at different spatio-temporal resolutions.
When using \urbannyc (Figure~\ref{fig:nyc_index}), the large
increase in time when moving from 3 to 4 \datasets was due to the
4$^{\mathrm{th}}$ \dataset, the Taxi data, which is not only large but
also contains many attributes. It also has the highest resolution both
in space and time, requiring each scalar function to be computed over
different resolutions.
There was also a significant increase in the number of computations
when the Weather \dataset was introduced (the 8$^{\mathrm{th}}$
\dataset): this \dataset has 228 numerical attributes.  However,
since it is relatively small compared to other \datasets in \urbannyc,
the running time was not significantly affected.

Given the large number of \datasets in \opendata, the total time taken
to identify the features was significantly larger than for \urbannyc (Figure~\ref{fig:open_index}).
Unlike \urbannyc, relatively more time is spent on feature identification
than on computing the scalar functions. 
This is due to two reasons: (1)~the \datasets in this collection are
much smaller; 
(2)~most of the 3D \datasets in \opendata are already in zip code
resolution, making it faster to compute the scalar functions; in
contrast, tuples in the 3D \urbannyc \datasets are GPS points,
which required additional computations to aggregate into the
neighborhood or zip code resolutions.
The total time taken to compute the 
indexes and 
features for 
all \datasets in \urbannyc and \opendata
was about 1 hour and 4 hours, respectively.

\begin{figure}[t!]
\centering
\subfigure[\urbannyc]{\label{fig:nyc_query}
  \includegraphics[height=0.13\textheight]{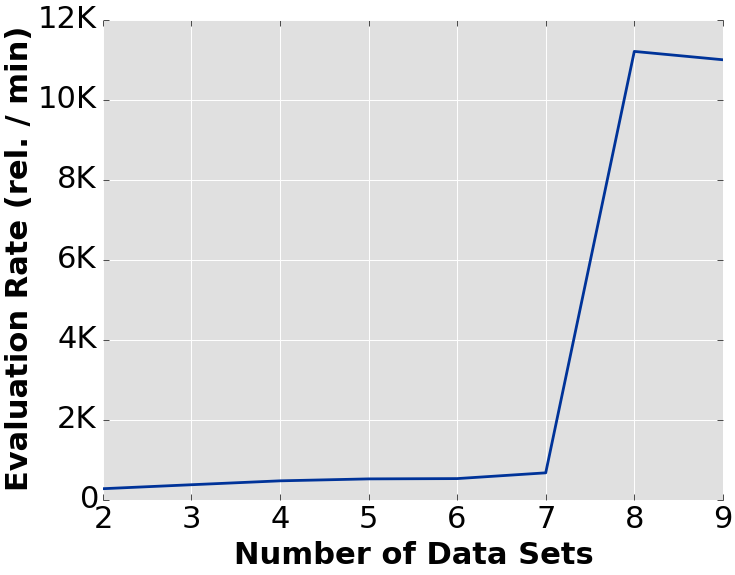}}~
\subfigure[\opendata]{\label{fig:open_query}
  \includegraphics[height=0.13\textheight]{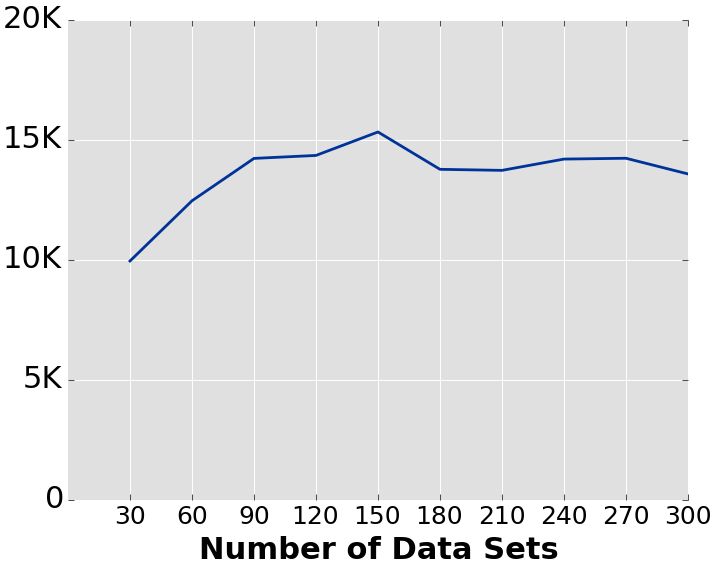}}~
\vspace{-0.2in}
\caption{Query performance.}
\label{fig:query}
\vspace{-.3in}
\end{figure}

\paragraph{Query Performance} 
To test the efficiency of the querying component, we executed a series of
queries that identify the relationships between a fixed
number of \datasets. 
Figure~\ref{fig:query} plots the relationship evaluation rate
with increasing number of \datasets.
Using both collections, \urbannyc and \opendata, we were able to consistently
evaluate relationships at a rate greater than $10^4$ relationships
per minute. The evaluation rate stabilized once the
number of relationships increased above this number, \eg with the
addition of the Weather data (\dataset 8) in
Figure~\ref{fig:nyc_query}.
A total of 290~thousand relationships were  
evaluated when the query used all \datasets from \urbannyc; for the
\opendata, this number was 17.4 million.
\highlight{The constant rate, irrespective of the data set pairs, 
indicates that relationship evaluation is independent of size and resolution
of the original data. This can be attributed to the abstraction of the data
as functions.}
Note that over 90\% of the querying time is spent on the
statistical significance tests, which involve
re-evaluating each relationship for 1,000 random spatial or temporal
permutations.
Also, the queries executed did not include any filtering clause.
When using a clause $C$, the query
evaluation step skips the significance test when $C$
is not satisfied, which further improves the performance.

\setlength{\columnsep}{4pt}
\begin{wrapfigure}{r}{1.6in}
\centering
\vspace{-0.15in}
\includegraphics[width=1.6in]{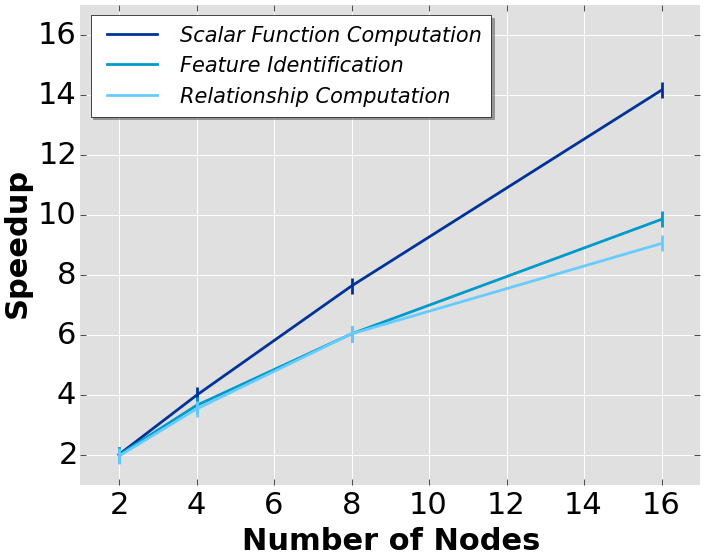}
\vspace{-0.3in}
\caption{Speedup}
\label{fig:speedup}
\vspace{-0.1in}
\end{wrapfigure} 
\paragraph{Scalability}
\highlight{To test the scalability of our framework, 
we computed the speedup attained by the different components with
increasing number of nodes in the cluster.
This experiment was performed on Amazon Web Services~(AWS) using the
\urbannyc collection.  We used AWS because it allows the configurations of clusters of
different sizes.
Each node in the cluster had an Intel Xeon E5-2670 v2 processor
(8-core) and 61GB of RAM.
Figure~\ref{fig:speedup} shows the speedup for the three components of the framework, which was computed against the 
the time taken by a single node. 
A relatively lower speedup was attained for identifying features and
evaluating relationships than for the scalar function computation.
This is primarily due to the presence of straggler reducers
that deal with higher spatio-temporal resolutions, thus increasing the
computation time for the randomization tests.}
\setlength{\columnsep}{\oldcolumnsep}

\begin{figure}[t]
\centering
\subfigure[\urbannyc]{\label{fig:nyc_week_city}
  \includegraphics[height=0.13\textheight]{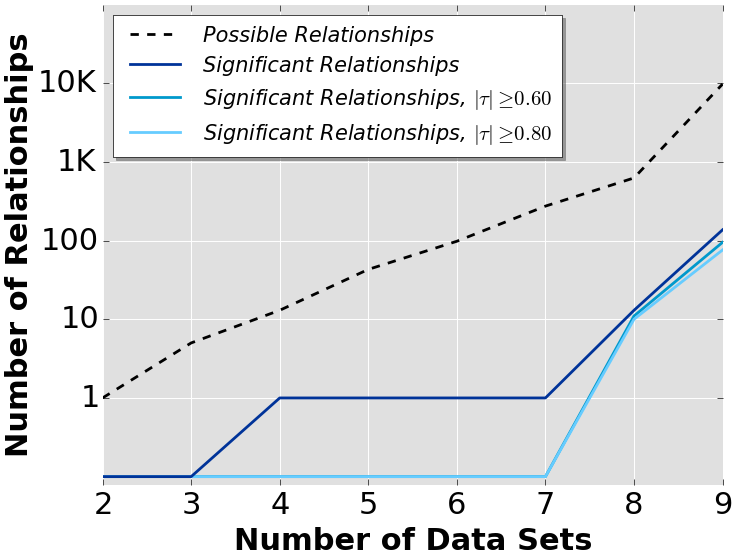}}~
\subfigure[\opendata]{\label{fig:open_week_city}
  \includegraphics[height=0.13\textheight]{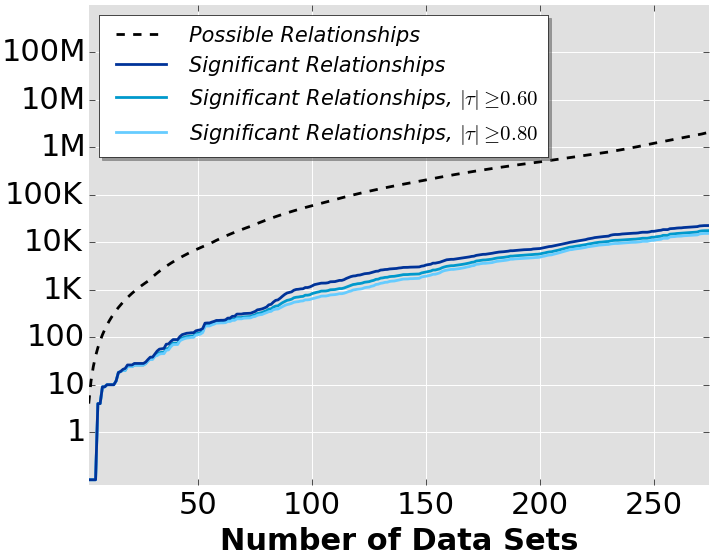}}
\vspace{-0.2in}
\caption{
Relationship pruning.
}
\label{fig:sparsity}
\vspace{-0.2in}
\end{figure}

\paragraph{Relationship Pruning}
Figure~\ref{fig:sparsity} plots the the number of 
identified relationships (in log scale), when considering the (week, city)  resolution,
with increasing number of \datasets.
For the \datasets in \urbannyc (Figure~\ref{fig:nyc_week_city}),
there was a significant
decrease in the number of relationships---from 9,745 to 137, a
decrease of about 98.60\%. When we filtered relationships having
$\tau \ge 0.6$ and $\tau \ge 0.8$, the reduction further increased to
99\% and 99.20\%, respectively.
When handling a larger number of \datasets, such as \opendata, the
advantages of our framework become even more evident, as 
Figure~\ref{fig:open_week_city} shows.
Given the over 2 million possible relationships for the (week, city)
resolution, our framework identified 22,327 of them to be
statistically significant, which corresponds to a decrease of about
98.90\%. Although the number of identified relationships is still
large, this is significantly better than trying to make sense of over
2 million relations. 
In addition, we envision that users will explore these relationships
by searching, querying, and filtering them based on different
attributes (\eg $\tau$, $\rho$, $\alpha$, space, and time).

\vspace{-0.2cm}
\subsection{Correctness and Robustness}
\label{sec:syn-evaluation}
\vspace{-0.1cm}

\paragraph{Correctness}
Most urban \datasets have become available only recently and work on
integrating them is still incipient~\cite{barbosa@bigdata2014}.
Since \emph{there are no ground-truth benchmarks} that can be used 
to evaluate the correctness of the identified relationships,
we used prior knowledge about the \datasets and designed a controlled
experiment to test if our technique can uncover strong relationships
that are expected to occur.
Consider the Taxi data in Figure~\ref{fig:motivation}. The density functions for
taxi trips in 2011 and 2012 have a similar behavior: the number of
trips in the city over time follows a similar pattern over the
two years, except in specific situations, such as during extreme
weather conditions.
Thus, if each year of data is modeled as a function (starting at the
same day and time), a strong positive relationship should be
observed for the two functions.
This observation was used to test our technique, which indeed
identified the two functions to be strongly and significantly related
across different resolutions.  The relationship score and strength for
the (hour, city) and the (hour, neighborhood) resolutions were
($\tau=0.99$, $\rho=0.85$) and ($\tau=1$, $\rho=0.87$),
respectively.

\begin{figure}[t]
\centering
\subfigure[Score]{\label{fig:robust_score}
  \includegraphics[height=0.13\textheight]{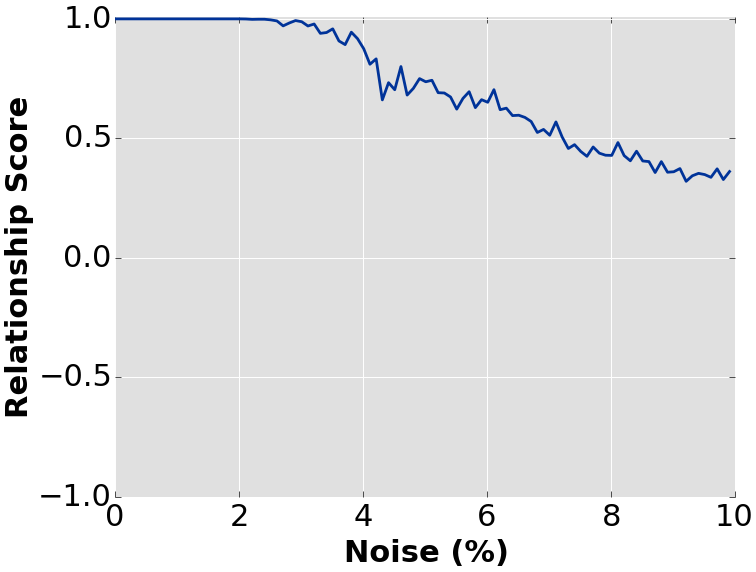}}~
\subfigure[Strength]{\label{fig:robust_strength}
  \includegraphics[height=0.13\textheight]{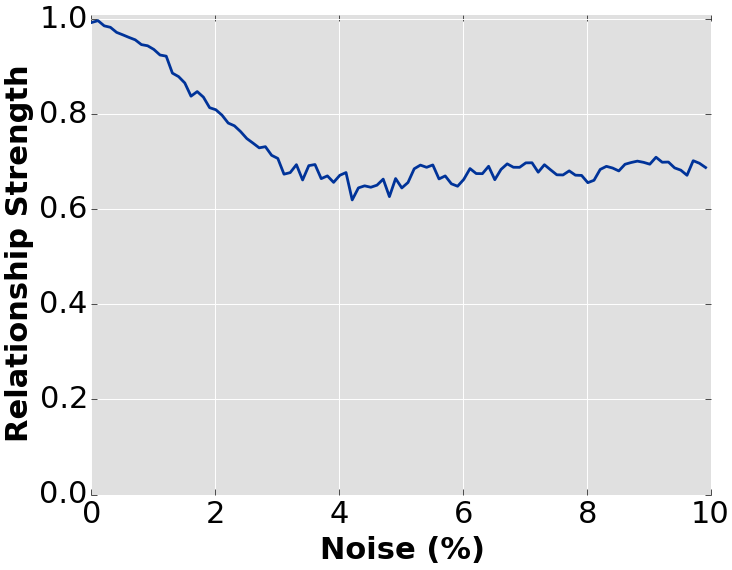}}
\vspace{-0.2in}
\caption{Robustness evaluation using the density of taxi trips.
}
\label{fig:robustness}
\vspace{-.3in}
\end{figure}

\paragraph{Robustness}
To assess the robustness of our technique in the presence of noise, we
fixed a scalar function $f$, and by artificially introducing noise to
$f$, we created a new (noisy) function $f^*$. We used a random
Gaussian noise where the amount of noise was bounded by a fraction of
the inter-quartile range of the function. Note that noise was added to
every spatio-temporal point of the function domain.
We then evaluated the relationship between $f$ and
$f^*$. Figure~\ref{fig:robustness} plots the relationship scores and
strengths with increasing levels of noise added to the taxi density
function. Note that even when the added noise was as large as $10\%$ of
the normal function range, we were still able to obtain a strong
positive relationship (which was statistically significant)
between the two functions. Furthermore, the relationship score remained
1 even when the noise level was greater than $2\%$.
This behavior can be attributed to the fact that topological
persistence, which is used to identify thresholds for the salient features, is robust
to noise~\cite{persistencediag}; small local
maxima and minima, which are created due to the addition of noise, do not
significantly affect the feature threshold. 

\vspace{-0.2cm}
\subsection{Effectiveness: Interesting Relationships}
\label{sec:qual-evaluation}

We carried out a detailed study using \urbannyc to assess the
effectiveness of our approach at finding interesting relationships and
pruning uninformative relationships that are not statistically
significant.
In our evaluation, we found Weather to be the most
\textit{polygamous} \dataset,
being related through
different attributes with all \datasets in the collection, except Gas
Prices, indicating the impact it has on different aspects of a city.
We discuss some of these relationships below (additional relationships
are described in Appendix~\ref{app:appendix-relationships}).
Unless otherwise noted, all relationships are with respect to salient features.
Note that, while some of the results might imply the presence of a causal
relationship, further analysis (not in the scope of this work) is
required to ascertain causality.

\paragraph{Weather and Taxi}
One of the relationships identified for the Taxi and Weather \datasets in the (hour, city) resolution
is between the 
number of taxis and the average precipitation, having score
$\tau=-0.62$ and strength $\rho=0.75$.
The values indicate a strong negative relationship: the higher the precipitation, the lower the
number of taxis in the city. As discussed in Section~\ref{sec:intro}, this confirms the
difficulty in finding taxis on a rainy day.
When testing the hypothesis that this is due to taxi drivers being target earners, 
we found a positive relationship between average fare and precipitation 
($\tau=0.73,\rho=0.7$) implying increased earnings when it rains. 
Note that Farber~\cite{farber@nber2014} refuted this hypothesis since he 
did not find a correlation (using OLS regression) between the drivers' earnings and rainfall. 
This is primarily due to two reasons: (i)~he did not take into account
the amount of rainfall---instead, he used a binary value indicating whether it rained or not;
and (ii)~more importantly, he considered the entire time period---periods
with very sparse rainfall are considered equivalent to  
those having higher rainfall. 
Thus, this case study also provides evidence for the importance of
looking at salient features,  since they can evince
relationships that are not visible when the whole data is taken into account.

While examining relationships involving extreme features, we found a
negative relationship between the number of trips and the average wind
speed. This relationship has a high score ($\tau = -1$) and a low
strength ($\rho=0.13$).  The low strength is due to the presence of
significant drops in the number of taxi trips at other periods, for
example, during Thanksgiving, Christmas, and New
Year~\cite{taxivis@tvcg2013} that are unrelated to the wind speed.
However, the high score indicates that, whenever there is high wind
speed, the number of taxi trips is significantly lower than usual,
which is related to the impact of hurricanes in the city.
We also found the same impact in the relationship
between number of unique taxis and average precipitation (having
$\tau=-1.0$) when considering extreme features.
It is worth noting that the number of taxi trips and wind speed are
not related through salient features alone (also observed from
Figure~\ref{fig:motivation}).  This demonstrates the importance of
computing relationships for salient as well as extreme features.

\paragraph{Weather and Citi Bike}
We found a positive
relationship between the average snow precipitation and the average
bike trip duration in the (hour, city) resolution, having $\tau=0.61$
and $\rho=0.16$.
This implies that bike trips are longer in snowy days (or shorter when
there is no snow), which is consistent with what we would expect.
For a lower resolution---(day, city)---we found a negative
relationship between the average snow precipitation and the active
Citi Bike stations ($\tau=-0.88$ and $\rho=0.65$), \ie fewer bike
stations are used when it snows. We believe that this is related not
only to the drop in bike usage under such weather, but also because
heavier snow may impact certain stations more than others: the city
clears snow at different frequencies depending on the location, and
some stations may get cleared faster than others.
Note that the latter relationship had a low score ($\tau=0$) when
considered at the higher resolution (hour, city):
the impact is usually reflected only after the snow accumulates,
and such accumulation is not captured when using an hourly time step (higher resolution). 
This case illustrates the need for evaluating relationships at
multiple resolutions.

\paragraph{Vehicle Collisions and Weather}
We found interesting relationships between Vehicle Collisions
and Weather, which correspond to the increased danger of
accidents when it rains. There was a strong positive relationship between
rainfall and number of motorists killed ($\tau=0.90,\rho=0.95$) as
well as number of injured pedestrians ($\tau=0.75,\rho=0.66$).
However, we found no significant relationship between number of
accidents and rainfall, implying that, even though the number of
accidents does not increase when there is heavy rain, their severity
does.
This leads to a new hypothesis that may explain the lack of taxis during rainfall:
taxi drivers, being experienced with the possible danger during high rainfall,
might return home during these periods.

\paragraph{Taxi and Traffic Speed}
We found a positive relationship between the average taxi fare 
and the average traffic speed at the (hour, neighborhood) resolution with $\tau=0.79$ and $\rho=0.44$,
implying that drivers may in fact earn less in the presence of heavy traffic. 
We also found strong negative relationships between the number of
taxi trips and the average traffic speed in the (hour, city) 
resolution ($\tau=-0.90$ and $\rho=0.65$).
This is expected, especially in a city such as NYC, which has around 13,000 taxis:
a larger number of trips increases the number of cars, thus slowing down the traffic.

\paragraph{Vehicle Collisions, 311, and Taxi}
We identified relationships that involve the
Vehicle Collision \dataset at the (hour, neighborhood) resolution:
a strong positive relationship between the number of
collisions and the number of 311 complaints ($\tau=0.99,\rho=0.86$),
and a strong positive relationship between the number
of collisions and number of taxi trips ($\tau=0.99$, $\rho=0.79$).
While the implication of such relationships, especially the latter, is
not clear, it provides a starting point for experts, pointing them
towards \datasets and attributes to be considered for further detailed
analysis.

\paragraph{Effectiveness of Statistical Significance Test}
Since there is no gold data available, to test the ability of the
significance tests to remove potentially uninformative relationships,
we evaluated a randomly chosen set of relationships that are not
statistically significant.
For instance, many relationships between the fare tax for taxi trips and
different attributes from the Weather, 311, and 911 \datasets were found not to be statistically
significant. This indicates that, even though some of these relationship have $|\tau|>0.60$,
they are mostly random and coincidental. In fact, the tax charged in taxi trip fares
does not have anything to do with different weather conditions, let alone with
311 and 911 complaints.
Other examples of spurious relationships pruned by our framework
include: mileage of taxi trips (Taxi) and number of injured
pedestrians (Vehicle Collisions), having $\tau=0.90$; number of bike
trips (Citi Bike) and number of tweets (Twitter), having $\tau=0.87$;
and number of 311 complaints and average speed (Traffic Speed), having
$\tau=0.76$.

We also computed the statistical significance for relevant
relationships using the standard Monte Carlo procedure.
Many of these relationships were found to be not significant
using this test, including the ones between the average snow
precipitation and the average bike trip duration. This underscores the
importance of taking the spatial and temporal dependencies into
account while assessing statistical significance.

Note that such tests represent a best-effort approach to identify
candidate relationships, and thus, they can both return spurious
relationships and miss important ones. 
Gold data are needed to quantitatively study the trade-offs for the
different techniques.
Nonetheless, our initial
experiments indicate that these significance tests are useful and can
help guide users in the data discovery process. 

\highlight{
\vspace{-.2cm}
\subsection{Comparison against Standard Techniques}
\label{sec:baseline-comparison}

We used the \urbannyc collection to compare our approach against established techniques for
identifying dependencies between data: Pearson
correlation coefficient (PCC) \cite{statistics}, mutual
information (MI) criterion \cite{su@hpdc2014}, and dynamic time
warping (DTW) routines \cite{keogh@2004,sakoe1978dynamic}.
Some of these techniques are not naturally normalized for
inter-dataset comparisons (\eg DTW and MI) or directly extendable to
spatio-temporal data. 
Thus, for this experiment, we proposed normalizations that provide a
meaningful range of relationship score 
(see Appendix~\ref{app:baseline-scores} for details) and focused on data
represented as a time series aggregated over the city resolution.

Overall, we observed that standard approaches can identify the basic
relationships that are present across the entire data.
For example, the relationship between average snow precipitation and
Citi Bike trip duration could be detected by PCC as well as by
MI. Similarly, the relationship between number of taxi trips and
average traffic speed could be found using PCC and DTW.
However, these techniques did not find relationships that are only
visible under certain conditions, such as the ones between rainfall
and number of taxis, or wind speed and number of taxi
trips. 
Also, relationships that take into account space, such as the
ones between number of collisions and number of taxi trips, are not
identified by any of the above techniques due to their inherent 1D
nature.
We would like to note that these approaches are orthogonal to our topology-based technique. 
Used in conjunction with our framework, we believe it will open new avenues in data discovery.

}

\vspace{-0.2cm}
\section{Related Work}
\label{sec:related}

In addition to the approaches discussed in Section~\ref{sec:baseline-comparison},
other notions of relationship
have also been explored  by the data mining and data integration
communities.
Some methods focused on identifying relationships between data points 
\emph{within a single} \dataset. 
Achtert~et~al.~\cite{achtert@sdm2007} focused on computing correlation clusters,
which are composed of data points that present correlations between different attributes.
Yang~et~al.~\cite{yang@icde2002} used subspace clustering, which finds
different clusters of points for different subspaces of attributes, to identify relationships.
Other methods have been proposed which identify different kinds of
relationships. Sarma et al.~\cite{sarma@sigmod2012} focused on finding
candidate tables that can be unioned and joined, and considered such
tables to be related.
There has also been work on \emph{data fusion}, where relationships
are sought between \datasets that overlap or complement each other to
resolve conflicts between different sources~\cite{dong@vldb2012,pochampally@sigmod2014}
and to find their derivation history~\cite{alawini@ssdbm2014}.
These relationships are orthogonal to and can be used in conjunction
with our technique to enrich the data discovery process.
To the best of our knowledge, no existing method addresses the problem
of identifying spatio-temporal relationships that take into account
salient features in the data. 

Recently, there has been a renewed interest in finding explanations for
surprising results, or \emph{outliers}, in database queries~\cite{roy@sigmod2014,wu@vldb2013}.
Scorpion~\cite{wu@vldb2013} focused on understanding aggregate queries over  a single \dataset.
Roy and Suciu~\cite{roy@sigmod2014}  handled more complex
database schemas and proposed techniques that can be used to explain relationships between different \datasets.
However, the thresholds for outliers must be specified by the user for each query,
which becomes impractical when handling hundreds to thousands
attributes 
and without
proper knowledge about the \datasets.
Since our \datapolygamy framework generates an overview of the
relationships among different \datasets, the approach proposed by Roy
and Suciu~\cite{roy@sigmod2014} could be used to further explore and
understand the most eye-catching relationships. 
Thus, the techniques complement each other in the data exploration process.

Methods for comparing scalar functions
use topological abstractions directly for this comparison (\eg \cite{Bauer2014, NTN2015}),
due to which 
two functions are considered to be
similar even if some affine transformation of the functions are
similar, \ie spatio-temporal locations of the topological features are not considered.
Unlike such methods, we are interested in comparing scalar functions
based on the spatio-temporal locations of their topological features.

\vspace{-0.1in}
\section{Discussion and Future Work}
\label{sec:conclusion}

\paragraph{Scalar Functions}
In this work, we mainly considered data whose spatial domain has dimension up to two. 
However, our framework is general and can handle higher dimensions
as well. For instance, data corresponding to noise in buildings can be obtained in 3D,
where in addition to geo-location, the noise level varies with height. By constructing an
appropriate graph to represent this spatial domain, the framework can be used as is. 

While we have focused on numerical attributes, non-numerical
attributes can be taken into account if they are mapped into
numerical values (\eg categorical values can be mapped to unique
numbers).
In addition, while we chose to use the average to represent functions,
it is straightforward to extend our framework to support other
functions such as \emph{sum}, \emph{median}, \emph{min}, or \emph{max}.
Alternatively, 
users can define custom
functions as well.

\highlight{
\paragraph{Types of Features}
Our current feature identification approach can miss unusual patterns
in the data due to its use of a single threshold.  For example, a
sudden increase in taxi trips in a relatively calm area and time will
not be identified if the density of taxi trips does not exceed the
computed threshold.  Instead, we could consider the \emph{gradient} of
this density over space and time.  High values of this function
correspond to regions that have sudden increase/decrease of function
value: the increase during a non-busy hour will show as a
high-gradient region, and can thus be identified as a feature.

In future work, we plan to do a comprehensive study of the different
types of scalar functions that can be derived from urban data sets,
and use them to classify the types of features they identify.  We
will use this classification to create a taxonomy of scalar functions
which can then be used by domain experts to appropriately choose the
types of features and relationships in which they are interested.
We will also investigate the use of a ROC curve which considers
multiple thresholds. This would help users pick the operating point of
interest depending on the desired sensitivity-specificity trade-off.}

\paragraph{Spatio-Temporal Resolution}
We currently support conversion from one resolution to another only when they
are compatible.
We plan to explore methods to support conversion between resolutions that do not
have a direct translation (\eg neighborhood and zip code) in order to
help evaluate relationships directly at a higher resolution
rather than moving to a lower resolution
(the latter may result in loss of information).

\paragraph{Future Extensions}
In this paper, we used salient, topological features as the basis for
identifying spatio-temporal relationships across disparate
\datasets. One direction we would like to explore is the use of event
detection techniques as an alternative to the topological features.
While the topology-based approach identifies local features corresponding to maxima and minima, event detection techniques must first create a model for what constitutes normal behavior to detect events that do not follow this behavior. Even though this can be expensive computationally, especially for spatio-temporal data, it would be interesting to study the performance trade-offs of the two approaches as well as compare the quality of the relationships they derive.

\highlight{
With the help of domain experts, we intend to use the
  available open data to create a benchmark for evaluating existing
  and future techniques.
While we are able to prune a significant number of relationships, a large number of
them might still need to be explored. We plan to design 
a visual interface to help this exploration process.}
We also plan to explore techniques to identify relationships that are
causal.
Finally, we intend to extend the statistical significance test to use
a 3-torus to incorporate both space and time together.

\paragraph{Acknowledgments} We thank Divesh Srivastava for his
feedback on an early draft of this paper and the anonymous SIGMOD
reviewers for their insightful comments and suggestions. This work was
supported in part by NSF awards CNS-1229185 and CI-EN-1405927, and by
the Moore-Sloan Data Science Environment at NYU. Juliana Freire is
partially supported by the DARPA Memex program, and Theodoros Damoulas
is partially supported by Alan Turing Institute.

\balance

{
\small
\putbib[paper,topology,machlearn]
}

\end{bibunit}

\begin{bibunit}[abbrv]
\appendix
\renewcommand{\thefigure}{\Roman{figure}}
\setcounter{figure}{0} 

\section{Table of Symbols}
\label{app:appendix-symbol}

\begin{table}[h]
\centerline{
\begin{tabular}{ | c | l |}
\hline
\cellcolor{black!20!white!50}\textbf{Symbol}
  & \multicolumn{1}{c|}{\cellcolor{black!20!white!50}\textbf{Description}} \\
\hline
$\Dspace$ & \Dataset
 \\
\hline
$A$, $B$ & \Dataset attribute
  \\
\hline
$f$, $g$ & Time-varying scalar function
  \\
\hline
$\Sspace$ & Spatial domain
  \\
\hline
$\Tspace$ & Temporal domain
  \\
\hline
$\theta$ & Feature threshold
  \\
\hline
$\theta^+$ & Threshold for positive features
  \\
\hline
$\theta^-$ & Threshold for negative features
  \\
\hline
$\Sigma_i$ & Set of features of a scalar function
  \\
\hline
$\Sigma$ & Set of feature-related points
  \\
\hline
$\Sigma_i^+$ & Set of positive features of a scalar function
  \\
\hline
$\Sigma_i^-$ & Set of negative features of a scalar function
  \\
\hline
$G$	& Graph representing the spatio-temporal domain \\
    & of a scalar function \\
\hline
$\tau$ & Relationship score
  \\
\hline
$\rho$ & Relationship strength
  \\
\hline
\end{tabular}}
\label{table:symbols}
\end{table}

\section{Additional Notes on Topology}
\label{app:topology}

\subsection{Critical Points and Morse Functions}
\label{app:morse}

Given a smooth, real-valued function $f:~\Rspace^d~\rightarrow~\Rspace$, a critical point $c_p$ of $f$ is \emph{non-degenerate} if the determinant of the \textit{Hessian} matrix at $c_p$ is non-singular.
The function $f$ is called a \emph{Morse function} if it satisfies the following conditions~\cite{CEHN04}:
\begin{enumerate}
\item All critical points of $f$ are non-degenerate.
\item All critical values are distinct \emph{i.e.}, $f(p) \neq f(q)$ for all critical points $p \neq q$.
\end{enumerate}
An important property of Morse functions is that critical points of such a function can be classified based on the behavior of the function within a local neighborhood~\cite{morsebook}.
In case of a PL function defined on a graph, the local neighborhood of a vertex $v$ is defined using the \emph{link} of that vertex.

\begin{definition}
The \emph{link} of a vertex $v$ is the sub-graph induced by the vertices adjacent on $v$. The \emph{upper link} of $v$ is the sub-graph induced by adjacent vertices having function value greater than $v$, while the \emph{lower link} of $v$ is the sub-graph induced by adjacent vertices having function value lower than $v$.
\end{definition}

Banchoff~\cite{Ban70} and Edelsbrunner et al.~\cite{EHNP03} describe a combinatorial characterization for critical points of a PL function, which are always located at vertices of the graph. 
Critical points are characterized by the number of connected components of the lower and upper links.
The vertex is \emph{regular} if it has exactly one lower link component and one upper link component. All other vertices are \emph{critical}. A critical point is a \emph{maximum} if the upper link is empty and a \emph{minimum} if the lower link is empty. 
It is a \textit{simple saddle} if it has one upper link and two lower link components, or two upper link and one lower link components. 
All other critical points are degenerate.

Both the conditions for a Morse function typically do not hold in practice for PL functions. 
In case of a degenerate critical point, it can be split into multiple simple saddles as shown by Edelsbrunner et al.~\cite{EHNP03} and Carr et al.~\cite{CSA03}.
A simulated perturbation of the function~\cite{Ede01} ensures that no two critical values are equal. This perturbation is accomplished by adding an infinitesimally small value to the vertices such that it imposes a total order on the vertices of the graph domain. This helps in consistently identifying the vertex with the higher function value between a pair of vertices, thus ensuring that condition~2 holds.
Note that in the latter case, when considering a flat local minimum (or maximum), one point in that region is identified as critical. This does not affect the identified features, since the thresholds are computed based on the persistence of the critical points, which does not change because of the location~\cite{persistencediag}.

\subsection{Merge Tree Computation}
\label{app:merge-tree}
For completeness, we now briefly describe the algorithm to compute merge trees of a PL function. For a more detailed description, we refer the reader to the work by Carr~et~al.~\cite{CSA03}. 
The algorithm uses the local neighborhood of a vertex in $G$ to classify whether the vertex is a critical point or not. The merge tree is built based on this classification. 

The join tree is computed by first sorting the vertices of $G$ in decreasing order of function value. Next, for each vertex $v$ in this sorted list, the algorithm performs the following operations: 
\begin{itemize}[leftmargin=*] \denselist
\item If $v$ is a maximum, create a new component containing $v$ and set $v$ as its \emph{head}. A vertex is a maximum if its upper link is empty~\cite{Ban70}. This is a direct extension of Definition~\ref{def:cp} to PL functions.
\item If the vertices in the upper link belong to one component, then the vertex is not critical. Add $v$ to this component.
\item $v$ is a critical point and not a maximum. Given a Morse function, the upper link in this case consists of 2 components~\cite{CSA03,EHNP03}. Add an edge between $v$ and the head of each of the 2 components. Next, merge these components and set $v$ as the head of the merged component. 
\end{itemize}
Similarly, the split tree is computed by traversing the vertices of $G$ in increasing order of function values and looking at the lower link of the vertices. The main operations performed in the algorithm are: (i)~creating components; (ii)~looking up components; and (iii)~merging components. This can be efficiently accomplished using the union-find data structure~\cite{CLR01}.

\paragraph{Time Complexity}
Let $N = |V|$ be the number of vertices in $G$. When computing the join / split trees, 
the vertices are first sorted, which takes $O(N \log N)$ time.
Then, for each vertex $v$, a \emph{create component} or \emph{merge component} operation is called once,
resulting in a total of $N$ such operations. 
A component look up of a vertex $v$ is executed whenever a vertex in its upper link is processed.
So, the total number of look up operations is equal to $|E| / 2$ (total degree over all vertices).
The spatial domains considered in this work correspond to cities, which are planar. Thus, $|E| = O(N)$.
Therefore, there is a total of $O(N)$ union-find operations requiring $O(N \alpha(N))$ time, where $\alpha()$ is the 
inverse Ackermann function. 
Combining all these steps, the join and split trees can be computed in $O(N \log N + N\alpha(N))$ time.

\section{Implementation}
\label{app:implementation}

The number of scalar functions to be computed is proportional to the
product of the total number of attributes in the data collection and
the different resolutions.
For the urban data sets we assessed (Section~\ref{sec:exp}),
thousands of scalar functions were computed, and the number of
relationships evaluated during querying was in the order of
millions. 
However, both the indexing and querying operations can be run independently for each
scalar function and scalar function pair, respectively.
To leverage the 
\emph{embarrassingly parallel} nature of these computations, we implemented the framework using \mapreduce.
We use three \mapreduce jobs: 

\begin{itemize}[leftmargin=*] \denselist
\item \emph{Scalar Function Computation Job.}  The map phase of the
  scalar function computation job maps each data point (or tuple) into
  the appropriate spatio-temporal points based on the resolution. The
  reduce phase aggregates data from each spatio-temporal point and
  generates all the scalar function values for that point.
\item \emph{Feature Identification Job.}  The map phase splits the
  different scalar functions based on the spatio-temporal resolution,
  while the reduce phase creates the merge tree index and identifies the
  features for a single function.  Recall that each spatio-temporal
  resolution of a (\dataset, attribute) pair is represented by one
  function.  
\item \emph{Relationship Computation Job.}  The map phase generates
  the different possible combinations between \datasets based on the
  input query (including different resolutions and feature types).  The
  reduce phase then evaluates each pair of functions based on the
  pre-computed set of features.  To evaluate a relationship, we need to
  compute the intersection of feature sets.  This is accomplished
  by representing each set of features as a bit vector, which not only reduces the
  memory size, but also allows us to use efficient bit operations to
  compute this intersection.
\end{itemize}

We use Hadoop for the \mapreduce implementation
and HDFS as the distributed file system, where the input \datasets,
intermediate data, and output are stored.
We use compression (BZip2 or Snappy codecs) for both map and reduce outputs,
and this led to significant speed ups, in particular for large
\datasets,  as fewer bytes need to be read and written.
We also use caching to speed-up future queries: when a query is
issued, the framework first checks if it can be answered using the
cache before invoking the \mapreduce job.
In case \textsc{clause} provides a threshold with respect to a \dataset, then features must 
first be identified before evaluating the relationship.

\highlight{
\section{Standard Approaches}
\label{app:baseline-scores}

In Section~\ref{sec:baseline-comparison}, we compared our approach against
three well-known techniques for identifying correlations:
Pearson correlation coefficient (PCC) \cite{statistics},
mutual information (MI) \cite{su@hpdc2014}, and
dynamic time warping (DTW) \cite{keogh@2004,sakoe1978dynamic}.
In what follows, we describe these approaches in detail.

\paragraph{Pearson's Correlation Coefficient (PCC)}
The Pearson's Correlation Coefficient
measures the linear correlation between two variables~\cite{statistics}.
Let $X$ and $Y$ be discrete random variables.
The PCC score $\beta_{PCC}$ can be computed as:
\begin{equation*}
\beta_{PCC}(X,Y) = \rho_{X,Y} = \frac{\mathrm{cov}(X,Y)}{\sigma_X\sigma_Y}
\end{equation*}
where \emph{cov} is the covariance and $\sigma$ is the
standard deviation.
The PCC score is a value between -1 and +1, where -1
indicates total negative correlation, +1 indicates
total positive correlation, and 0 indicates
no linear correlation.

\paragraph{Mutual Information (MI)}
In information theory, mutual information quantifies the amount of information 
obtained about one random variable through another random variable. Formally, 
the MI between two discrete random variables $X$ and $Y$ is given as:
\begin{equation*}
I(X,Y) = \sum_{i \in X}\sum_{j \in Y} P(i,j)\mathrm{log}\frac{P(i,j)}{P(i)P(y)}
\end{equation*}
where $P(i,j)$ is the joint probability distribution function of $X$ and $Y$, 
and $P(i)$ and $P(j)$ are marginal distribution functions of $X$ and $Y$,
respectively.
Since we wanted to compare different pairs of variables, we used a
normalized variant of MI ($\beta_{MI}$):
\begin{equation*}
\beta_{MI}(X,Y) = \frac{I(X,Y)}{\sqrt{H(X)H(Y)}}
\end{equation*}
where $H(X)$ and $H(Y)$ are the Shannon entropies of $X$ and $Y$, respectively.
The MI score ranges from 0 to 1: while 0 indicates that the variables are
independent, 1 specifies that variables are completely dependent.

\paragraph{Dynamic Time Warping (DTW)}
Dynamic Time Warping is a dynamic programming approach
\cite{sakoe1978dynamic,salvador@2007} that is used to construct a
\textit{minimum edit-distance} between two temporal sequences.
Traditionally, DTW has been used to compare
different pairs of time series and perform clustering.
However, since DTW distances are not naturally normalized,
they cannot be directly compared across multiple time series.
Therefore, we propose and employ the following normalized DTW score $\beta_{DTW}$:
\begin{equation*}
\beta_{DTW}(X,Y) = 1 - \frac{\mathrm{DTW}(X,Y)}{\mathrm{DTW}(X,0)+\mathrm{DTW}(0,Y)}
\end{equation*}
where $0$ represents a constant line at $y=0$,
and variables $X$ and $Y$ are Z-normalized.
This score ranges from 0 (dissimilar/uncorrelated) to 1 (identical/correlated) and allowed us to compare across DTW scores and time series distances.

}

\section{Additional Experiments}
\label{app:experiments}

\subsection{Robustness Evaluation}

\begin{figure}[t]
\centering
\subfigure[Score]{\label{fig:robust_score_unique}
  \includegraphics[height=0.13\textheight]{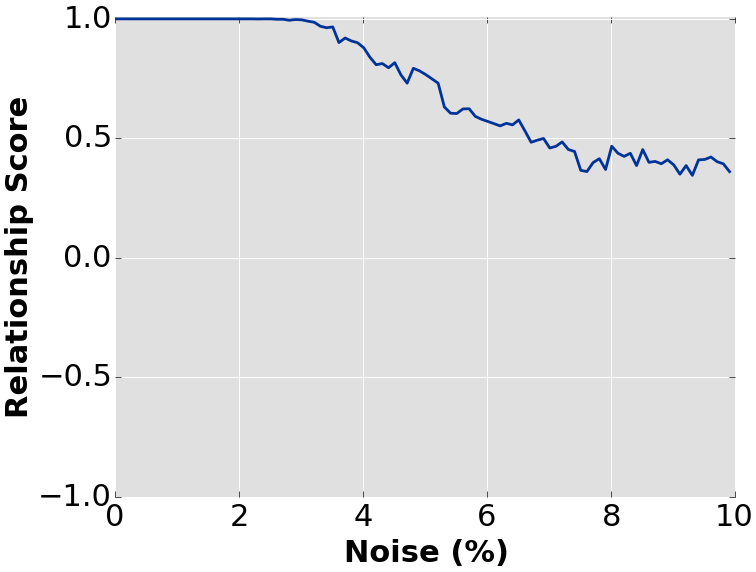}}~
\subfigure[Strength]{\label{fig:robust_strength_unique}
  \includegraphics[height=0.13\textheight]{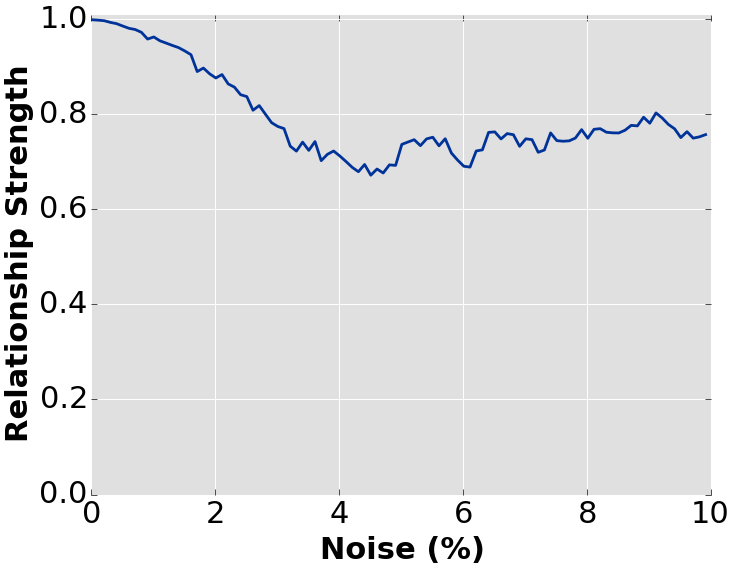}}
\vspace{-0.2in}
\caption{Robustness evaluation (unique number of taxis).
}
\label{fig:robustness-unique}

\subfigure[Score]{\label{fig:robust_score_avg_miles}
  \includegraphics[height=0.13\textheight]{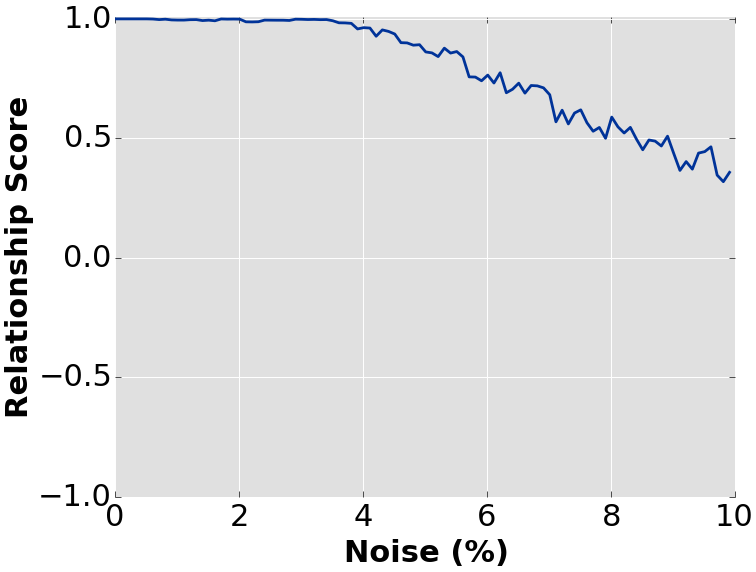}}~
\subfigure[Strength]{\label{fig:robust_strength_avg_miles}
  \includegraphics[height=0.13\textheight]{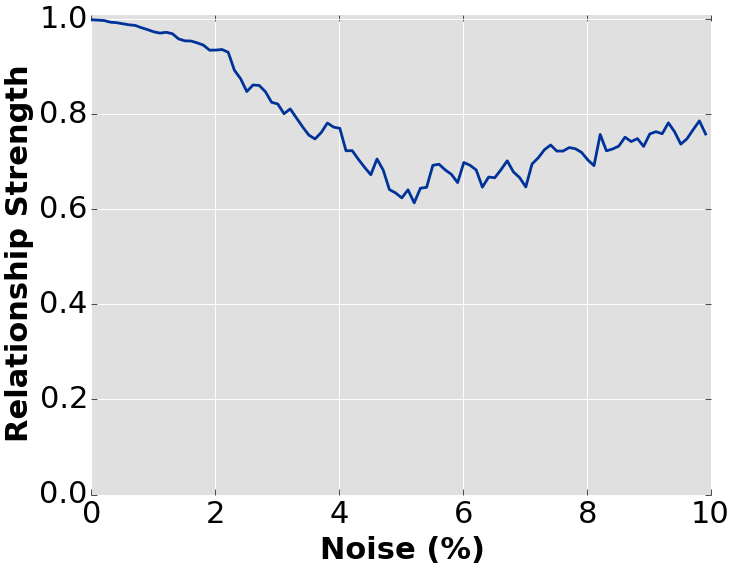}}
\vspace{-0.2in}
\caption{Robustness evaluation (average of traveled miles).
}
\label{fig:robustness-avg-miles}

\subfigure[Score]{\label{fig:robust_score_avg_fare}
  \includegraphics[height=0.13\textheight]{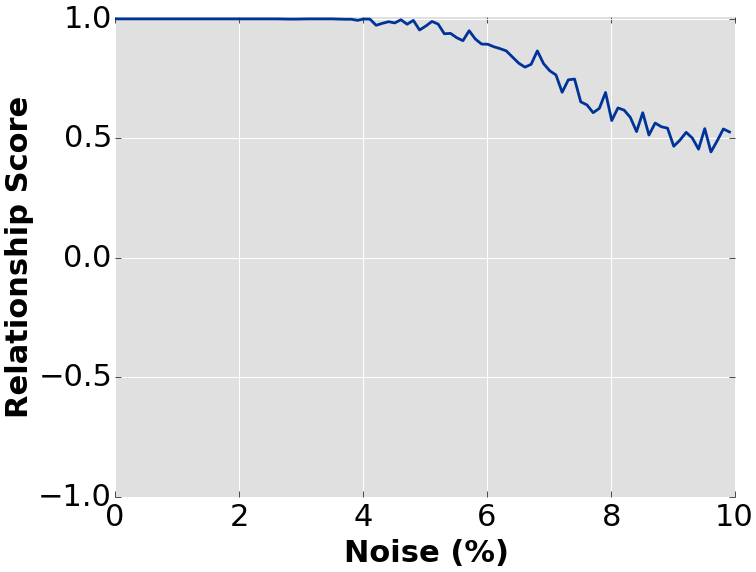}}~
\subfigure[Strength]{\label{fig:robust_strength_avg_fare}
  \includegraphics[height=0.13\textheight]{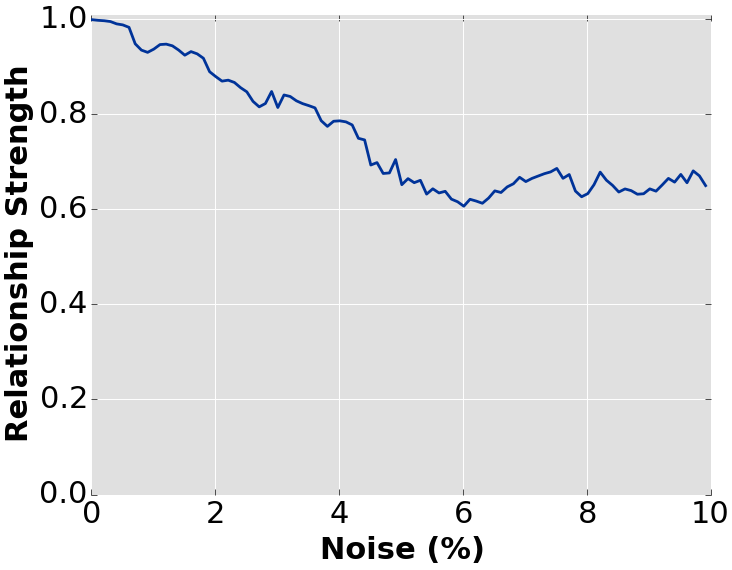}}
\vspace{-0.2in}
\caption{Robustness evaluation (average of total taxi fare).
}
\label{fig:robustness-avg-fare}
\vspace{-0.15in}
\end{figure}

In addition to the taxi density function, we also tested the robustness of our
technique for other scalar functions of the Taxi \dataset, 
including the unique function (Figure~\ref{fig:robustness-unique}), 
the average miles function (Figure~\ref{fig:robustness-avg-miles}),
and the average fare function (Figure~\ref{fig:robustness-avg-fare}).
Similar to the result reported in Section~\ref{sec:syn-evaluation} for the
density function, we find that our technique is robust with other
functions as well.

\subsection{Interesting Relationships}
\label{app:appendix-relationships}

In this section, we discuss some additional interesting and 
potentially informative relationships that were identified
for the \urbannyc collection.

\paragraph{Weather and Taxi}
A relationship between the unique number of taxi medallions and the average precipitation could also be identified in the (day, city) resolution with a higher score than in the (hour, city) resolution: $\tau=-0.81$.
We also found out that the unique number of taxi medallions is positively related to the average visibility in the (day, city) resolution ($\tau=0.7,\rho=0.45$),
and negatively related to the average snow depth in the (week, city) resolution ($\tau=-0.95,\rho=0.95$): this might indicate that some taxi drivers avoid working when the weather conditions are not appropriate (\eg low visibility or high snow accumulation in streets).
There was also a negative relationship between the number of taxi trips and the average snow precipitation in the (week, city) resolution ($\tau=-0.87,\rho=0.82$) and in the (month, city) resolution ($\tau=-1.0,\rho=0.07$).

\paragraph{Weather and Citi Bike}
We identified negative relationships that involve the unique number of Citi Bikes with the average snow precipitation ($\tau=-1.0,\rho=0.19$),
the average snow depth ($\tau=-0.62,\rho=0.45$),
and rainfall ($\tau=-0.62,\rho=0.44$), all in the (day, city) resolution. This indicates that less Citi Bikes are used when weather conditions are unpleasant. 

\paragraph{Weather and Traffic Speed}
For the Weather and Traffic Speed \datasets, we found a positive relationship between the average visibility and the average traffic speed in the (week, city) resolution ($\tau=0.75,\rho=0.24$) and in the (month, city) resolution ($\tau=1.0,\rho=0.14$).
Such relationships are expected: when the visibility in the streets is low (\eg due to a foggy weather), cars tend to drive slower to avoid accidents.

\paragraph{Taxi and Traffic Speed}
Recall that we discovered a negative relationship between the number of taxi 
trips and the average traffic speed in the (hour, city) resolution. 
This relationship was also identified in other resolutions, including 
(day, neighborhood) ($\tau=-0.99,\rho=0.08$); (day, city) ($\tau=-0.84,\rho=0.4$); and (week, city) ($\tau=-1.0,\rho=0.16$).
Additionally, there was also a negative relationship between the average taxi trip duration and the average traffic speed in the (day, city) resolution, having $\tau=-0.93$ and $\rho=0.51$. 

\paragraph{Taxi and Gas Prices}
We found a positive relationship between the average fare and the average gas price in the (month, city) resolution, with $\tau=1.0$ and $\rho=0.5$: in fact, the per-mile metered taxi rate may increase when gas prices also increase~\cite{dailynews-taxi}. We find this at a monthly resolution because it is difficult to detect the affect of gas prices at a lower resolution. Note that the Gas Prices data is available at a weekly resolution.
We also found a negative relationship between the unique number of taxis and the average gas price in (week, city) resolution ($\tau=-1.0,\rho=0.008$) and (month, city) resolution ($\tau=-1.0,\rho=0.5$).
However, the implication of this relationship is not clear.

\paragraph{311 and 911}
There were some curious relationships between the number of 311 complaints
and the number of 911 calls, in both (day, neighborhood) resolution ($\tau=0.92,\rho=0.27$) and (week, neighborhood) resolution ($\tau=1.0,\rho=0.65$).
Experts can perform further analysis to better understand these relationships and find out if emergency and non-emergency calls are indeed related.

\paragraph{Vehicle Collisions}
We identified some interesting relationships between Vehicle Collisions and other \datasets. In the (day, city) resolution, there was a positive relationship between the average number of motorists injured in a vehicle crash and the average traffic speed ($\tau=0.64,\rho=0.46$),
which might indicate that high speeds influence the occurrence of
collisions that are dangerous for drivers.
In the (day, neighborhood) resolution, there were relationships between the number of collisions and the number of 311 complaints ($\tau=0.84,\rho=0.41$),
and also with the number of 911 calls ($\tau=0.94,\rho=0.18$).
The former also showed up in the (week, neighborhood) resolution, with $\tau=0.72$ and $\rho=0.22$.
In this resolution, there was also a positive relationship between the number of collisions and the number of taxi trips ($\tau=0.99,\rho=0.25$).
Again, although these relationships may not have apparent implications, experts can use this information to perform further analysis and identify the reason behind them.

{
\small
\putbib[paper,topology,machlearn]
}
\end{bibunit}

\end{document}